\documentclass[
superscriptaddress,
showpacs,
amsmath,
amssymb,
aps,
prb,
twocolumn,
]{revtex4}

\usepackage{graphicx}
\usepackage{dcolumn}
\usepackage{bm}
\usepackage{hyperref}

\begin{document}


\title{Robustness of the Insulating Bulk in the Topological Kondo Insulator SmB$_{6}$}

\author{Y. S. Eo}
	\email{eohyung@umich.edu}
	\affiliation{University of Michigan, Dept.~of Physics, Ann Arbor, Michigan 48109-1040, USA}
\author{A. Rakoski}
	\affiliation{University of Michigan, Dept.~of Physics, Ann Arbor, Michigan 48109-1040, USA}
\author{J. Lucien}
	\affiliation{University of Michigan, Dept.~of Physics, Ann Arbor, Michigan 48109-1040, USA}
\author{D. Mihaliov}
	\affiliation{University of Michigan, Dept.~of Physics, Ann Arbor, Michigan 48109-1040, USA}
\author{\c{C}. Kurdak}  
	\affiliation{University of Michigan, Dept.~of Physics, Ann Arbor, Michigan 48109-1040, USA}
\author{P. F. S. Rosa}
	\affiliation{Condensed Matter $\&$ Magnet Science Group, Los Alamos National Laboratory, Los Alamos, NM 87545}	
\author{D.-J. Kim}
	\affiliation{University of California at Irvine, Dept.~of Physics and Astronomy, Irvine, California 92697, USA}
\author{Z. Fisk}
	\affiliation{University of California at Irvine, Dept.~of Physics and Astronomy, Irvine, California 92697, USA}
 
\date{\today}

\begin{abstract}

We used the inverted resistance method to extend the bulk resistivity of SmB$_{6}$ to a regime where the surface conduction overwhelms the bulk. Remarkably, the bulk resistivity shows an intrinsic thermally activated behavior that changes ten orders of magnitude, suggesting that it is an ideal insulator that is immune to disorder. Non-stoichiometrically-grown SmB$_{6}$ samples also show an almost identical thermally activated behavior. At low temperatures, however, these samples show a mysterious high bulk resistivity plateau, which may arise from extended defect conduction in a 3D TI.

\end{abstract}

\pacs{72.10.Bg, 71.10.Fk}%

\maketitle

 Semiconductors, or narrow band-gapped insulators, have been one of the most important classes of materials both for technological advances in electronics and for fundamental scientific studies in the past several decades. In technology, the realization of the modern electrical and optoelectrical devices that we use today are possible because of the successful control of point defects (donors and acceptors) in semiconductors. In fundamental science studies, semiconductors have provided a fascinating playground for the discovery of new states of matter. One example is 3D topological insulators (TIs)\cite{FuPRL,JEMoorePRB}, discovered about a decade ago\cite{DHsieh,YXiaNatPhys,Chen2009}, in which a unique two-dimensional electron gas emerges on the surface due to bulk band inversion. Many of the 3D TIs, however, can easily be found in the degenerate semiconductor regime or even in the hopping conduction regime where they exhibit large bulk conduction due to the presence of unintentional impurities and disorder\cite{BRAHLEK201554}. Obtaining a higher quality bulk in such 3D TIs is an on-going technical challenge\cite{BRAHLEK201554}. 
 
 Recently, there has been further excitement about the prediction of 3D TIs in strongly correlated insulators\cite{DzeroPRL, Takimoto}. The best candidate is SmB$_{6}$, a traditionally well-known mixed-valent insulator or Kondo insulator that has a narrow bulk band gap\cite{JWAllen1979}. In the bulk of SmB$_{6}$, an almost flat 4$f$  band and a dispersive 5$d$ band hybridize at cryogenic temperatures forming a very small band gap at the Fermi level. Theory suggests that this hybridization plays the role of band-inversion, resulting in a 3D TI\cite{DzeroPRL, Takimoto}. Electrical transport experiments unambiguously revealed the existence of the conducting surface at low temperatures (below 3-4 K), consistent with the 3D TI prediction\cite{WolgastPRB,kim2013surface}. Below 3-4 K, the insulating bulk becomes too resistive for the current to flow in, and the current flows on the conducting surface instead, revealing itself as a resistance plateau. Despite some of the earlier angle-resolved photoemission spectroscopy (ARPES) results suggesting that the conducting surface is from a trivial origin \cite{zhu2013polarity}, most recent experimental works including electrical\cite{WolgastPRB,kim2013surface,wolgast2015magnetotransport,nakajima2016one,syers2015tuning,thomas2016weak,PhysRevB.91.085107} and thermal transport\cite{luo2015heavy}, de Haas-van Alphen quantum oscillations by angle-dependent magneto-torque magnetometry\cite{li2014two}, ARPES\cite{neupane2013surface,denlinger2014smb6,jiang2013observation,NXU1,NXU2}, scanning tunneling microscopy\cite{yee2013imaging,rossler2014hybridization,miyamachi2017evidence}, point contact spectroscopy\cite{zhang2013hybridization}, planar tunneling spectroscopy\cite{park2016topological,sun2017planar}, and neutron scattering\cite{fuhrman2015interaction} have provided strong evidence of the existence of conducting surface states that is consistent with the unique properties of a 3D TI. There are even reports of the helical spin-structure, which is one of the most unique properties of a topologically protected surface, by spin-resolved ARPES measurements\cite{xu2014direct} and a recent report of a spin-signal on the surface by the inverse Edelstein effect\cite{song2016spin}.
 
 The quantum oscillations by magneto-torque measurements have provided the research community with some of the most exciting yet confusing results in this material system. G. Li $et$ $al$.\cite{li2014two} report on quantum oscillations indicating the presence of 2D Fermi pockets on the (100) and (110) surfaces. Furthermore, the extrapolation of the Landau indices from the oscillations to the infinite magnetic field limit reveals a Berry phase contribution that is consistent with the Dirac-like dispersion that emerges in a 3D TI. In contrast, B. S. Tan $et$ $al$. \cite{tan2015unconventional} later report on quantum oscillations with a non-2D angle dependence, deviation of the standard Lifshitz-Kosevich temperature dependence, a Berry phase that is non-Dirac-like, and oscillation amplitudes that do not depend on the surface facets. They conclude that their oscillations originate from an unconventional Fermi surface in the insulating bulk. The possibility of having a material with a bulk Fermi surface in the absence of a conducting Fermi liquid resulted in further excitement about SmB$_{6}$. To explain the unconventional Fermi surface of the bulk, new theories have been developed involving exotic excitations that couple to the magnetic field but not the electric field\cite{PhysRevLett.119.057603,PhysRevLett.118.096604}. 
 
 Vibrant research in the past several years has been motivated by both the need of verifying the 3D TI properties and the search for new exotic bulk phenomena of SmB$_{6}$. One aspect, however, that remains yet elusive is the role of disorder. For example, some of the early studies of SmB$_{6}$ report that samples with vacancies result in a lower resistance plateau\cite{kasuya1977theory} that is inconsistent with the 3D TI picture. This is inconsistent because higher vacancy levels are expected to introduce higher disorder on the surface, and therefore they should have higher surface resistivity. Also to the best of our knowledge, many recent experimental conclusions and theoretical predictions, including the interpretations of the quantum oscillation reports, assume that the SmB$_{6}$ crystal is close to ideal. We believe it is critically important to consider the role of disorder in the bulk in more depth before the community moves on to discuss new exciting intrinsic properties of the SmB$_{6}$ crystals. 
 
 In this letter, we use transport measurements on SmB$_{6}$, including stoichiometrically- and non-stoichiometrically-grown samples, to study the role of disorder in the bulk. We first note that the characterization of bulk transport in the presence of significant surface conduction is challenging. We have previously argued that the common practice of presenting the residual-resistance ratio from conventional four-contact resistance measurements is limited when interpreting a material system that has both surface and bulk conducting states\cite{eo2017new}. In this study, we use the inverted resistance measurement technique, which was proposed by the authors recently\cite{eo2017new}. Inverted resistance measurements performed on multi-ring Corbino structures allows us to properly characterize the bulk conduction even in the presence of strong surface conduction. We demonstrate the substantial difference when we choose a typical four-contact measurement compared to a proper Corbino disk measurement in Supplementary A.
 
\begin{table}[h]
\squeezetable
\centering 
\begin{tabular}{c c c c c} 
\hline\hline 
Sample & $r$& $E_a$ (meV) & $\rho_{s}$ (k$\Omega$) & Hardness(kp) \\ [1ex]
\hline 
S1 & 0 & 4.01&3.1 & 2191 $\pm$ 125 \\ 
S2 & 0.1 & 4.12&1.6 & 1913 $\pm$ 16.5 \\
S3 & 0.25 & 3.97&1.5	 & 1781 $\pm$ 111 \\
S4 & 0.40 & 3.85&2.8 & 1563 $\pm$ 50.8\\[1ex] 
\hline 
\end{tabular}
\caption{Summary of characterization of the crystals. Samples were grown with a starting composition of ratio, $r$, where Sm:B:Al = 1-$r$: 6: 700. Detailed information of crystal characterization, including X-ray, Auger, and hardness measurements can be found in Supplementary B.}
\label{Tab:samples}
\end{table}
 
 Single crystalline samples were grown by the Al-flux technique. The mixture of Samarium pieces (Ames Lab, 99.99 $\%$), Boron powder (99.99 $\%$) and Aluminum shots (99.999 $\%$) was placed in an alumina crucible and loaded in a vertical tube furnace with ultra high-purity Ar flow. We grew samples with different starting compositions of Sm: B: Al = 1-$r$: 6: 700, where $r$ is the starting composition ratio, ranging from 0 to 0.40. We expect sample S1 in which $r$=0 to be stoichiometric, whereas the other samples are expected to have a higher disorder level. X-ray and Auger electron spectroscopy measurements did not have the resolution to unambiguously determine the point defect levels, but we do see differences that indicate the overall physical properties are changing from hardness measurements. A detailed description is presented in Supplementary B. 
\begin{figure}[t]
\begin{center}
\includegraphics[scale=0.7]{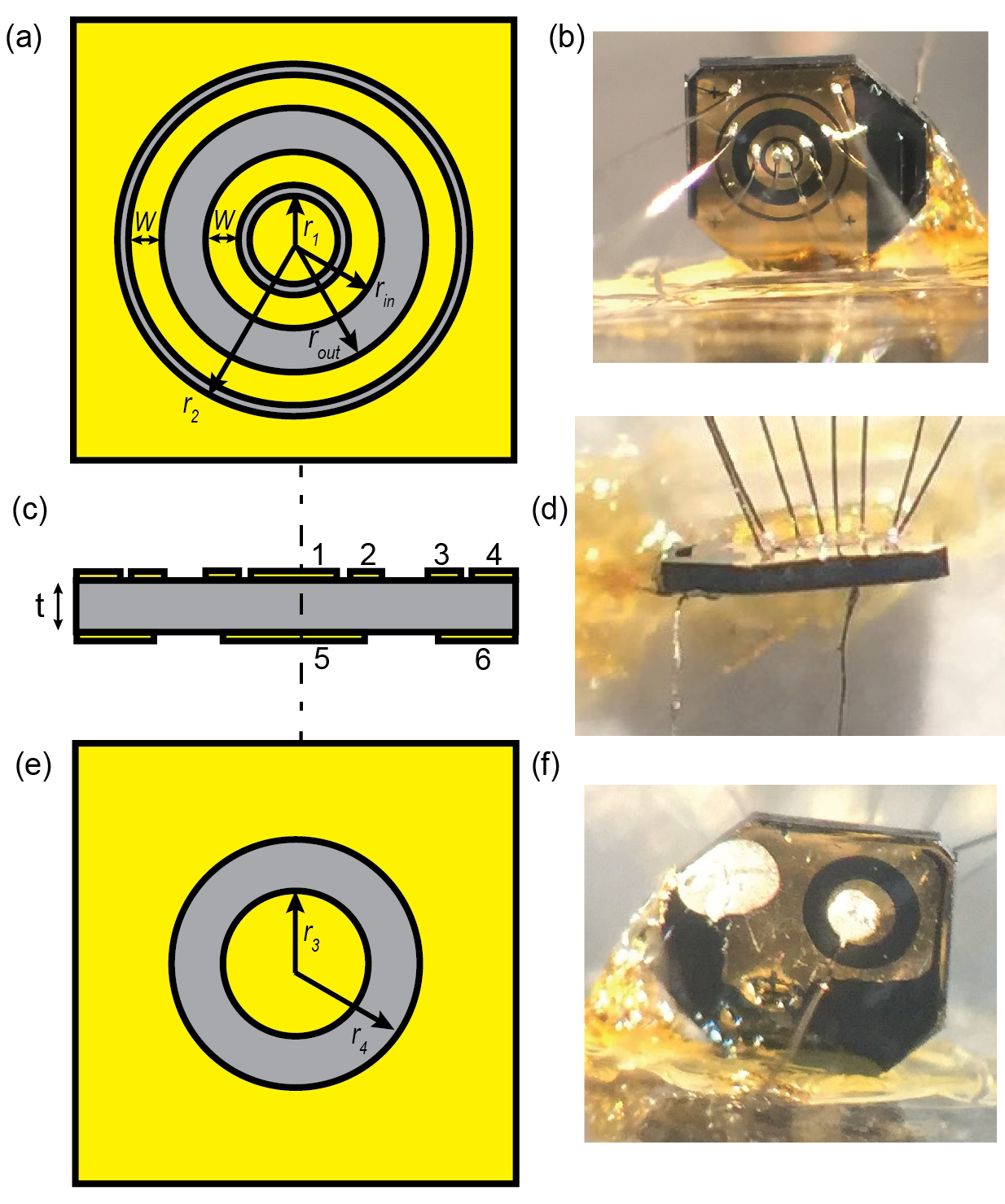}
\caption{(color) Schematic diagram and actual image of the transport geometry used in this experiment. Schematic diagram of (a) top surface, (c) side surface, and (e) bottom surface. Actual image of the (b) top surface, (d) side surface, and (f) bottom surface. The dimensions are: $r_{1}$ = 100 $\mathrm{\mu m}$, $r_{2}$ = 800 $\mathrm{\mu m}$, $r_{\mathrm{in}}$ = 200 $\mathrm{\mu m}$, $r_{\mathrm{out}}$ = 300 $\mathrm{\mu m}$, $W$ = 75 $\mathrm{\mu m}$, $r_{3}$ = 165 $\mathrm{\mu m}$, and $r_{4}$ = 290 $\mathrm{\mu m}$.}
\label{Fig:Schematic}
\end{center}
\end{figure}

 To illustrate the inverted resistance method in short, the transport geometry is shown in Fig.~\ref{Fig:Schematic}. The samples were fine polished with a final step of aluminum oxide slurry that has a particle size of 0.3 $\mu$m. The Corbino-disk patterns were fabricated using standard photolithography, followed by ebeam evaporation of Ti/Au (20\AA/1500\AA). We used a home-built instrumentation amplifier in addition to an external lock-in amplifier in the Dynacool PPMS for measurement. A four-terminal Corbino disk can be measured by $R_{1,4;2,3}$ (= $V_{2,3}$/$I_{1,4}$), which can be regarded as a standard resistance measurement ($R_{\mathrm{Std}}$). In the surface dominated regime, below 3-4 K, the inverted resistance ($R_{\mathrm{Inv}}$) can be measured by either $R_{1,4;5,6}$ or $R_{1,2;3,4}$. If the change in surface resistivity with temperature is not strong compared to the bulk, the standard two-channel model is a good approximation that works well for $R_{\mathrm{Std}}$ in the full temperature range:
\begin{equation}
	R_{\mathrm{Std}} = C_0 (\rho_{s}^{-1} + \gamma \rho_{b}^{-1})^{-1},
	\label{Eq:twochannel}
\end{equation}
where the geometric prefactor $C_0$ is ln$(r_{\mathrm{out}}/r_{\mathrm{in}})/2\pi$ for a Corbino disk, and $\gamma$ is the effective thickness that asympotically approaches $t$ when the sample is very thin, but is independent of $t$ when the sample is very thick\cite{eo2017new}. The inverted resistance below the bulk-to-surface crossover temperature follows:
\begin{equation}
	R_{\mathrm{Inv}} = C_1 t \frac{\rho_{s}^2}{\rho_{b}},
	\label{Eq:Inverted2}
\end{equation}
where $C_{1}$ is a prefactor for the inverted resistance. The corresponding $\gamma$ and $C_{1}$ are found from finite element analysis, similar to the derivation of bulk resistivity extraction in Ref.~\cite{eo2017new}. The detailed method is presented in Supplementary C. 
\begin{figure}[t]
\begin{center}
\includegraphics[scale=0.9]{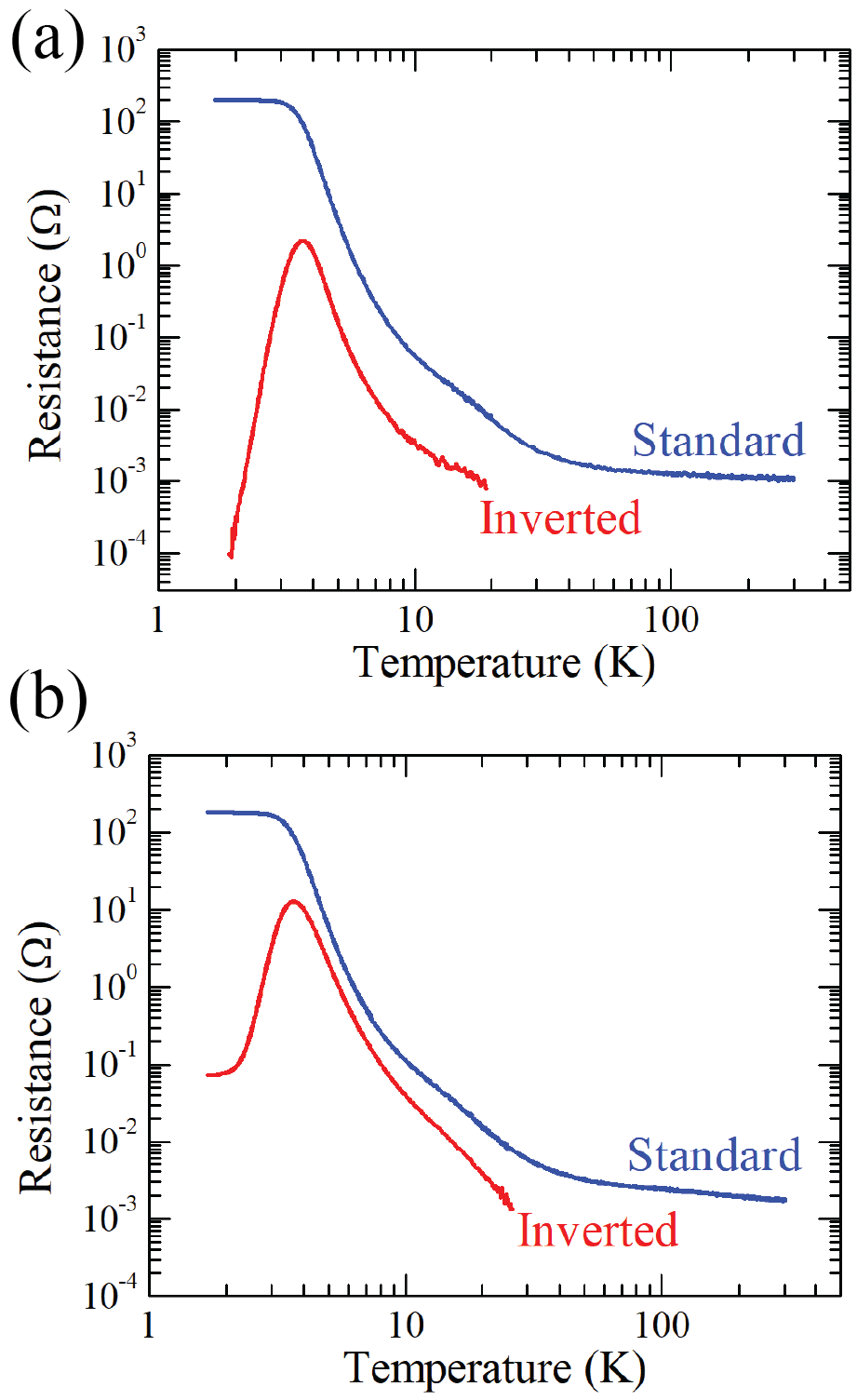}
\caption{(color) Resistance of the standard and inverted resistance measurement. (a) Results for the stoichiometrically-grown SmB$_{6}$ (sample S1).  (b) Results from sample S4.}
\label{Fig:ResistanceResults}
\end{center}
\end{figure}

 Fig.~\ref{Fig:ResistanceResults}~(a) shows the measured resistance ($R_{\mathrm{Std}}$ (blue) and $R_{\mathrm{Inv}}$ (red)) from sample S1. The qualitative behavior of Fig.~\ref{Fig:ResistanceResults}~(a) is consistent with what we expect when the bulk resistivity is governed intrinsically, $\rho_b \propto \mathrm{exp}(E_{a}/k_{B} T)$, where $E_a$ is the activation energy.  In the high-temperature regime, above $\sim$ 4 K, both $R_{\mathrm{Std}}$ and $R_{\mathrm{Inv}}$ increase when the temperature is lowered, consistent with Eq.~\ref{Eq:twochannel} in the bulk-dominated regime ($\rho_b / \rho_{s} t \rightarrow \infty$). Below $\sim$4 K, $R_{\mathrm{Std}}$ develops a plateau which corresponds to a sheet resistance of $\rho_s$ = 3 k$\Omega$ according to Eq.~\ref{Eq:twochannel} in the surface-dominated regime ($\rho_b / \rho_{s} t \rightarrow 0$). $R_{\mathrm{Inv}}$, on the other hand, drops as the temperature is lowered. This is consistent with Eq.~\ref{Eq:Inverted2} when the bulk resistivity follows $\rho_b \propto \mathrm{exp}(E_{a}/k_{B} T)$. Below 1.99 K, the inverted resistance becomes too small, and the measurement is limited by the amplifier performance. Here, we only present the data that is meaningful, above this performance limit.
 
 Next, we consider the non-stoichiometrically-grown SmB$_{6}$ samples. We present the sample S4 results in Fig.~\ref{Fig:ResistanceResults}~(b). In the bulk-dominated regime, above $\sim$ 4 K, the temperature response of the resistances of all samples behaved qualitatively identically to sample S1 results. In the surface-dominated regime, below $\sim$ 4 K, $R_{\mathrm{Std}}$ shows a plateau that corresponds to a $\rho_s$ in the 1 - 3 k$\Omega$ range. In contrast, $R_{Inv}$ in the surface-dominated regime drops at first, consistent with Eq.~\ref{Eq:Inverted2} when the bulk resistivity is keeps rising. Below $\sim$2.5 K, however, the resistance becomes much weakly dependent of temperature. The magnitude of this resistance plateau becomes lower for samples that are grown with less Sm. 
 \begin{figure}[t]
\begin{center}
\includegraphics[scale=0.9]{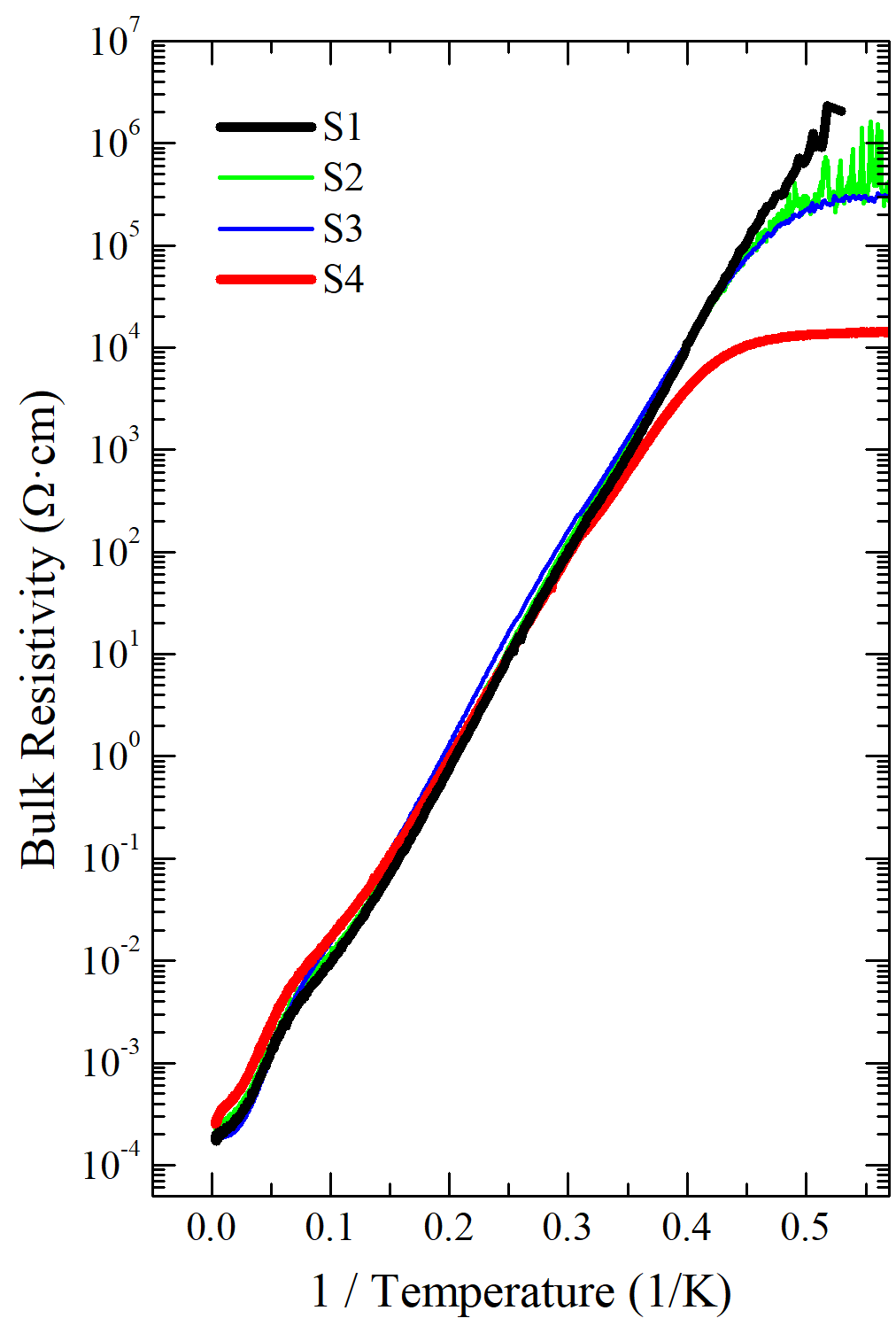}
\caption{(color) Bulk resistivity conversion of the stoichiometrically-grown (S1) and non-stoichiometrically-grown (S2, S3, and S4) SmB$_{6}$ samples. }
\label{Fig:ResistivityResults}
\end{center}
\end{figure}

 The bulk resistivities converted from the resistance measurements are shown in Fig.~\ref{Fig:ResistivityResults}. Sample S1 (shown in black) shows a thermally activated exponential behavior (E$_{a}$ = 4.01 meV) with a change of $\sim$10 orders of magnitude. The resistivity shows an intrinsic semiconductor behavior in the full temperature range, without showing any signs of the extrinsic regime of a semiconductor. Below 0.4 K$^{-1}$ (or above $\sim$2.5 K), the non-stoichiometrically-grown SmB$_6$ samples (Samples S2, S3, and S4) show almost identical activation energies, and the resistivity rises at least 7 orders of magnitude. 
 
 It is well known from previous Hall measurements that the bulk resistivity rise of SmB$_{6}$ is due to the thermally activated bulk carriers, and the mobility changes within two orders of magnitude \citep{JWAllen1979,molnar1982study,kasuya1983mechanisms,sluchanko1999nature,PhysRevB.61.9906}. In our stoichiometrically-grown sample (sample S1) result , the activation behavior continues at low temperatures indicating a decrease in carrier density of $\sim$10 orders of magnitude. We, therefore, estimate a carrier density of $10^{12}$ (1/cm$^3$) at 1.99 K. This implies that there are only $\sim 10^{7}$ carriers in the bulk region of our sample at this temperature, and there would be less than 1 bulk carrier below 1 K by extrapolating the thermally activated behavior. The non-stoichiometrically-grown samples (Samples S2, S3, and S4) also indicate that no more than $10^{15}$ (1/cm$^3$) bulk carriers are left before the bulk resistivity saturates below 2.5 K. For the saturation of the non-stoichiometrically-grown samples, the magnitudes are too high for them to be attributed to the extrinsic regime of a non-degenerate semiconductor when there are point defects. Note that the associated number of carriers of these resisitivity plateaus, of course, is too small to explain the exotic bulk quantum oscillations seen by Tan $et$ $al$\cite{tan2015unconventional}. Also, none of the data resembles close to the temperature dependence of the Mott-type variable range hopping conduction ($\rho \propto \mathrm{exp}(1/T)^{1/4}$). Recent AC conductivity measurements show large conduction that completely ruling out the impurity conduction scenario in the bulk is worrisome\cite{PhysRevB.94.165154,laurita2017impurities}. However, the conduction does not appear in the DC limit, consistent with our results. Our data show strong evidence that the bulk gap of SmB$_6$ does not involve point defect-like impurity states that can contribute to DC transport, and therefore the bulk is robust. It is important to note that this is in stark contrast to other hexaboride systems such as CaB$_{6}$\cite{stankiewicz2016electrical,rhyee2004effect,PhysRevB.62.10076}, SrB$_{6}$\cite{stankiewicz2016electrical,ott1997structure}, BaB$_{6}$\cite{stankiewicz2016electrical}, YbB$_{6}$\cite{kim2007weak, tarascon1980magnetic}, and EuB$_{6}$\cite{wigger2004electronic,kasuya1980iv}, which all show strong dependence of disorder in the temperature-dependent resistivity measurements. This result is also promising when comparing to the bulk of other 3D TIs, where purifying bulk is an ongoing challenge\cite{ando2013topological,BRAHLEK201554}.

 To the best of our knowledge, the robust intrinsic insulating behavior and the high bulk resistivity plateau in the non-stoichiometrically-grown samples cannot be explained by standard theory of disorder. We have previously reported that the picture of shallow impurity states, or in-gap states, in SmB$_6$ are not justified by the effective mass approximation framework because the effective Bohr radius, $a^*_B$, is comparable or smaller than the lattice constant\cite{PhysRevB.95.195133}. The small resisitivity feature around 14 K can be explained by the alternative model with a clean gap and band banding by the surface states\cite{PhysRevB.95.195133}. We find the most similar case, where the bulk gap appears to be uninterrupted by point defects, is the BCS gap in superconductors. It is well known that the BCS gap withstands dirty conditions, as long as the impurity is non-magnetic\cite{anderson1959theory}. 

 Within the framework that the bulk gap is robust against point defects, to understand the mysterious high bulk resistivity plateaus in the non-stoichiometrically-grown samples, it is most reasonable to think that the conduction path has a confined or lower-dimensional current channel instead of a homogeneous channel in the bulk. One plausible theory is that extended defects, such as grain boundaries and threading dislocations, conduct because SmB$_6$ can harbor topologically protected edge conduction. Particularly, the possibility of 1D dislocation conduction in a 3D TI is a further excitement, as it is a unique 1D edge conduction channel that is not localized by disorder\cite{ran2009one}. These need not be uniformly distributed, but instead exist sporadically with length scales that extend throughout the bulk of the sample. The cause of these high-order defects may be related to the inhomogeneity in the bulk. Our hardness and X-ray measurements, which are presented in the Supplementary B, show that all sample indeed have different physical properties, and signatures of disorder present in the crystal. In particular, in the single crystal X-ray diffraction measurement, we observe signatures of twinning in some of our samples. Lastly, it is worth to note that recent SmB$_{6}$ studies focus on the role of impurities. Heat capacity on vacancy and impurity doped SmB$_6$ show a virtual metallic-like behavior at low temperatures from the bulk\cite{valentine2017effect,fuhrman2017screened}, and recent thermal transport measurements show field-dependent thermal conductivity enhancement depending on the sample quality\cite{boulanger2017field}. We do not yet find a clear connection between these results and our bulk transport plateaus at low temperatures. For future studies, this mysterious bulk conduction channel must be studied in more depth, which include ongoing studies of other variously doped SmB$_6$ samples. 
 
In conclusion, the bulk transport of stoichiometrically- and non-stoichiometrically-grown SmB$_6$ samples were studied through the inverted resistance measurement. Using the double-sided Corbino disk geometry, the stoichiometrically-grown SmB$_6$ sample shows a robust thermally activated bulk resistivity rise. In the non-stoichiometrically-grown SmB$_6$ samples, grown with substantially less samarium, these samples show an almost identical thermally activated behavior until an unexpected resistivity plateau develops. Our results suggest that the bulk of SmB$_6$ is immune to disorder originating from point defects, but may be influenced by extended defects. We believe the robust insulating bulk in SmB$_6$ is important for TI applications. For example, in spintronics applications, there would be no parallel channel from the bulk. For topological quantum computers, because the lifetime of the Majorana modes will not be limited by the bulk channel, we expect the Majorana modes to be well defined.
\begin{acknowledgments}
We thank J. W. Allen, K. Sun, and J. Denlinger for useful discussions and advice on improving the manuscript. We also thank B. L. Scott for the discussion of the single crystal X-ray diffraction data, and Zhongrui(Jerry) Li for assistance with the Auger electron spectroscopy experiments. Funding for this work was provided by NSF Grants No. DMR1441965 and No. DMR-1643145. P. F. S. R acknowledges support from the Laboratory Directed Research and Development program of Los Alamos National Laboratory under project number 20160085DR. 
\end{acknowledgments} 

\bibliography{RobustBulkSmB6}

\begin{thebibliography}{59}%
\makeatletter
\providecommand \@ifxundefined [1]{%
 \@ifx{#1\undefined}
}%
\providecommand \@ifnum [1]{%
 \ifnum #1\expandafter \@firstoftwo
 \else \expandafter \@secondoftwo
 \fi
}%
\providecommand \@ifx [1]{%
 \ifx #1\expandafter \@firstoftwo
 \else \expandafter \@secondoftwo
 \fi
}%
\providecommand \natexlab [1]{#1}%
\providecommand \enquote  [1]{``#1''}%
\providecommand \bibnamefont  [1]{#1}%
\providecommand \bibfnamefont [1]{#1}%
\providecommand \citenamefont [1]{#1}%
\providecommand \href@noop [0]{\@secondoftwo}%
\providecommand \href [0]{\begingroup \@sanitize@url \@href}%
\providecommand \@href[1]{\@@startlink{#1}\@@href}%
\providecommand \@@href[1]{\endgroup#1\@@endlink}%
\providecommand \@sanitize@url [0]{\catcode `\\12\catcode `\$12\catcode
  `\&12\catcode `\#12\catcode `\^12\catcode `\_12\catcode `\%12\relax}%
\providecommand \@@startlink[1]{}%
\providecommand \@@endlink[0]{}%
\providecommand \url  [0]{\begingroup\@sanitize@url \@url }%
\providecommand \@url [1]{\endgroup\@href {#1}{\urlprefix }}%
\providecommand \urlprefix  [0]{URL }%
\providecommand \Eprint [0]{\href }%
\providecommand \doibase [0]{http://dx.doi.org/}%
\providecommand \selectlanguage [0]{\@gobble}%
\providecommand \bibinfo  [0]{\@secondoftwo}%
\providecommand \bibfield  [0]{\@secondoftwo}%
\providecommand \translation [1]{[#1]}%
\providecommand \BibitemOpen [0]{}%
\providecommand \bibitemStop [0]{}%
\providecommand \bibitemNoStop [0]{.\EOS\space}%
\providecommand \EOS [0]{\spacefactor3000\relax}%
\providecommand \BibitemShut  [1]{\csname bibitem#1\endcsname}%
\let\auto@bib@innerbib\@empty
\bibitem [{\citenamefont {Fu}\ \emph {et~al.}(2007)\citenamefont {Fu},
  \citenamefont {Kane},\ and\ \citenamefont {Mele}}]{FuPRL}%
  \BibitemOpen
  \bibfield  {author} {\bibinfo {author} {\bibfnamefont {L.}~\bibnamefont
  {Fu}}, \bibinfo {author} {\bibfnamefont {C.~L.}\ \bibnamefont {Kane}}, \ and\
  \bibinfo {author} {\bibfnamefont {E.~J.}\ \bibnamefont {Mele}},\ }\href
  {\doibase 10.1103/PhysRevLett.98.106803} {\bibfield  {journal} {\bibinfo
  {journal} {Phys. Rev. Lett.}\ }\textbf {\bibinfo {volume} {98}},\ \bibinfo
  {pages} {106803} (\bibinfo {year} {2007})}\BibitemShut {NoStop}%
\bibitem [{\citenamefont {Moore}\ and\ \citenamefont
  {Balents}(2007)}]{JEMoorePRB}%
  \BibitemOpen
  \bibfield  {author} {\bibinfo {author} {\bibfnamefont {J.~E.}\ \bibnamefont
  {Moore}}\ and\ \bibinfo {author} {\bibfnamefont {L.}~\bibnamefont
  {Balents}},\ }\href {\doibase 10.1103/PhysRevB.75.121306} {\bibfield
  {journal} {\bibinfo  {journal} {Phys. Rev. B}\ }\textbf {\bibinfo {volume}
  {75}},\ \bibinfo {pages} {121306} (\bibinfo {year} {2007})}\BibitemShut
  {NoStop}%
\bibitem [{\citenamefont {Hsieh}\ \emph {et~al.}(2008)\citenamefont {Hsieh},
  \citenamefont {Qian}, \citenamefont {Wray}, \citenamefont {Xia},
  \citenamefont {Hor}, \citenamefont {Cava},\ and\ \citenamefont
  {Hasan}}]{DHsieh}%
  \BibitemOpen
  \bibfield  {author} {\bibinfo {author} {\bibfnamefont {D.}~\bibnamefont
  {Hsieh}}, \bibinfo {author} {\bibfnamefont {D.}~\bibnamefont {Qian}},
  \bibinfo {author} {\bibfnamefont {L.}~\bibnamefont {Wray}}, \bibinfo {author}
  {\bibfnamefont {Y.}~\bibnamefont {Xia}}, \bibinfo {author} {\bibfnamefont
  {Y.}~\bibnamefont {Hor}}, \bibinfo {author} {\bibfnamefont {R.}~\bibnamefont
  {Cava}}, \ and\ \bibinfo {author} {\bibfnamefont {M.}~\bibnamefont {Hasan}},\
  }\href {http://www.nature.com/nature/journal/v452/n7190/abs/nature06843.html}
  {\bibfield  {journal} {\bibinfo  {journal} {Nature}\ }\textbf {\bibinfo
  {volume} {452}},\ \bibinfo {pages} {970} (\bibinfo {year}
  {2008})}\BibitemShut {NoStop}%
\bibitem [{\citenamefont {Xia}\ \emph {et~al.}(2009)\citenamefont {Xia},
  \citenamefont {Qian}, \citenamefont {Hsieh}, \citenamefont {Wray},
  \citenamefont {Pal}, \citenamefont {Lin}, \citenamefont {Bansil},
  \citenamefont {Grauer}, \citenamefont {Hor}, \citenamefont {Cava} \emph
  {et~al.}}]{YXiaNatPhys}%
  \BibitemOpen
  \bibfield  {author} {\bibinfo {author} {\bibfnamefont {Y.}~\bibnamefont
  {Xia}}, \bibinfo {author} {\bibfnamefont {D.}~\bibnamefont {Qian}}, \bibinfo
  {author} {\bibfnamefont {D.}~\bibnamefont {Hsieh}}, \bibinfo {author}
  {\bibfnamefont {L.}~\bibnamefont {Wray}}, \bibinfo {author} {\bibfnamefont
  {A.}~\bibnamefont {Pal}}, \bibinfo {author} {\bibfnamefont {H.}~\bibnamefont
  {Lin}}, \bibinfo {author} {\bibfnamefont {A.}~\bibnamefont {Bansil}},
  \bibinfo {author} {\bibfnamefont {D.}~\bibnamefont {Grauer}}, \bibinfo
  {author} {\bibfnamefont {Y.}~\bibnamefont {Hor}}, \bibinfo {author}
  {\bibfnamefont {R.}~\bibnamefont {Cava}},  \emph {et~al.},\ }\href
  {http://www.nature.com/nphys/journal/v5/n6/abs/nphys1274.html} {\bibfield
  {journal} {\bibinfo  {journal} {Nat. Phys.}\ }\textbf {\bibinfo {volume}
  {5}},\ \bibinfo {pages} {398} (\bibinfo {year} {2009})}\BibitemShut {NoStop}%
\bibitem [{\citenamefont {Chen}\ \emph {et~al.}(2009)\citenamefont {Chen},
  \citenamefont {Analytis}, \citenamefont {Chu}, \citenamefont {Liu},
  \citenamefont {Mo}, \citenamefont {Qi}, \citenamefont {Zhang}, \citenamefont
  {Lu}, \citenamefont {Dai}, \citenamefont {Fang} \emph {et~al.}}]{Chen2009}%
  \BibitemOpen
  \bibfield  {author} {\bibinfo {author} {\bibfnamefont {Y.}~\bibnamefont
  {Chen}}, \bibinfo {author} {\bibfnamefont {J.}~\bibnamefont {Analytis}},
  \bibinfo {author} {\bibfnamefont {J.-H.}\ \bibnamefont {Chu}}, \bibinfo
  {author} {\bibfnamefont {Z.}~\bibnamefont {Liu}}, \bibinfo {author}
  {\bibfnamefont {S.-K.}\ \bibnamefont {Mo}}, \bibinfo {author} {\bibfnamefont
  {X.-L.}\ \bibnamefont {Qi}}, \bibinfo {author} {\bibfnamefont
  {H.}~\bibnamefont {Zhang}}, \bibinfo {author} {\bibfnamefont
  {D.}~\bibnamefont {Lu}}, \bibinfo {author} {\bibfnamefont {X.}~\bibnamefont
  {Dai}}, \bibinfo {author} {\bibfnamefont {Z.}~\bibnamefont {Fang}},  \emph
  {et~al.},\ }\href {http://science.sciencemag.org/content/325/5937/178}
  {\bibfield  {journal} {\bibinfo  {journal} {Science}\ }\textbf {\bibinfo
  {volume} {325}},\ \bibinfo {pages} {178} (\bibinfo {year}
  {2009})}\BibitemShut {NoStop}%
\bibitem [{\citenamefont {Brahlek}\ \emph {et~al.}(2015)\citenamefont
  {Brahlek}, \citenamefont {Koirala}, \citenamefont {Bansal},\ and\
  \citenamefont {Oh}}]{BRAHLEK201554}%
  \BibitemOpen
  \bibfield  {author} {\bibinfo {author} {\bibfnamefont {M.}~\bibnamefont
  {Brahlek}}, \bibinfo {author} {\bibfnamefont {N.}~\bibnamefont {Koirala}},
  \bibinfo {author} {\bibfnamefont {N.}~\bibnamefont {Bansal}}, \ and\ \bibinfo
  {author} {\bibfnamefont {S.}~\bibnamefont {Oh}},\ }\href
  {https://www.sciencedirect.com/science/article/pii/S0038109814004426}
  {\bibfield  {journal} {\bibinfo  {journal} {Solid State Commun.}\ }\textbf
  {\bibinfo {volume} {215-216}},\ \bibinfo {pages} {54 } (\bibinfo {year}
  {2015})}\BibitemShut {NoStop}%
\bibitem [{\citenamefont {Dzero}\ \emph {et~al.}(2010)\citenamefont {Dzero},
  \citenamefont {Sun}, \citenamefont {Galitski},\ and\ \citenamefont
  {Coleman}}]{DzeroPRL}%
  \BibitemOpen
  \bibfield  {author} {\bibinfo {author} {\bibfnamefont {M.}~\bibnamefont
  {Dzero}}, \bibinfo {author} {\bibfnamefont {K.}~\bibnamefont {Sun}}, \bibinfo
  {author} {\bibfnamefont {V.}~\bibnamefont {Galitski}}, \ and\ \bibinfo
  {author} {\bibfnamefont {P.}~\bibnamefont {Coleman}},\ }\href {\doibase
  10.1103/PhysRevLett.104.106408} {\bibfield  {journal} {\bibinfo  {journal}
  {Phys. Rev. Lett.}\ }\textbf {\bibinfo {volume} {104}},\ \bibinfo {pages}
  {106408} (\bibinfo {year} {2010})}\BibitemShut {NoStop}%
\bibitem [{\citenamefont {Takimoto}(2011)}]{Takimoto}%
  \BibitemOpen
  \bibfield  {author} {\bibinfo {author} {\bibfnamefont {T.}~\bibnamefont
  {Takimoto}},\ }\href {http://journals.jps.jp/doi/abs/10.1143/JPSJ.80.123710}
  {\bibfield  {journal} {\bibinfo  {journal} {J. Phys. Soc. Jpn.}\ }\textbf
  {\bibinfo {volume} {80}},\ \bibinfo {pages} {123710} (\bibinfo {year}
  {2011})}\BibitemShut {NoStop}%
\bibitem [{\citenamefont {Allen}\ \emph {et~al.}(1979)\citenamefont {Allen},
  \citenamefont {Batlogg},\ and\ \citenamefont {Wachter}}]{JWAllen1979}%
  \BibitemOpen
  \bibfield  {author} {\bibinfo {author} {\bibfnamefont {J.~W.}\ \bibnamefont
  {Allen}}, \bibinfo {author} {\bibfnamefont {B.}~\bibnamefont {Batlogg}}, \
  and\ \bibinfo {author} {\bibfnamefont {P.}~\bibnamefont {Wachter}},\ }\href
  {\doibase 10.1103/PhysRevB.20.4807} {\bibfield  {journal} {\bibinfo
  {journal} {Phys. Rev. B}\ }\textbf {\bibinfo {volume} {20}},\ \bibinfo
  {pages} {4807} (\bibinfo {year} {1979})}\BibitemShut {NoStop}%
\bibitem [{\citenamefont {Wolgast}\ \emph {et~al.}(2013)\citenamefont
  {Wolgast}, \citenamefont {Kurdak}, \citenamefont {Sun}, \citenamefont
  {Allen}, \citenamefont {Kim},\ and\ \citenamefont {Fisk}}]{WolgastPRB}%
  \BibitemOpen
  \bibfield  {author} {\bibinfo {author} {\bibfnamefont {S.}~\bibnamefont
  {Wolgast}}, \bibinfo {author} {\bibfnamefont {C.}~\bibnamefont {Kurdak}},
  \bibinfo {author} {\bibfnamefont {K.}~\bibnamefont {Sun}}, \bibinfo {author}
  {\bibfnamefont {J.~W.}\ \bibnamefont {Allen}}, \bibinfo {author}
  {\bibfnamefont {D.-J.}\ \bibnamefont {Kim}}, \ and\ \bibinfo {author}
  {\bibfnamefont {Z.}~\bibnamefont {Fisk}},\ }\href {\doibase
  10.1103/PhysRevB.88.180405} {\bibfield  {journal} {\bibinfo  {journal} {Phys.
  Rev. B}\ }\textbf {\bibinfo {volume} {88}},\ \bibinfo {pages} {180405}
  (\bibinfo {year} {2013})}\BibitemShut {NoStop}%
\bibitem [{\citenamefont {Kim}\ \emph {et~al.}(2013)\citenamefont {Kim},
  \citenamefont {Thomas}, \citenamefont {Grant}, \citenamefont {Botimer},
  \citenamefont {Fisk},\ and\ \citenamefont {Xia}}]{kim2013surface}%
  \BibitemOpen
  \bibfield  {author} {\bibinfo {author} {\bibfnamefont {D.}~\bibnamefont
  {Kim}}, \bibinfo {author} {\bibfnamefont {S.}~\bibnamefont {Thomas}},
  \bibinfo {author} {\bibfnamefont {T.}~\bibnamefont {Grant}}, \bibinfo
  {author} {\bibfnamefont {J.}~\bibnamefont {Botimer}}, \bibinfo {author}
  {\bibfnamefont {Z.}~\bibnamefont {Fisk}}, \ and\ \bibinfo {author}
  {\bibfnamefont {J.}~\bibnamefont {Xia}},\ }\href
  {https://www.ncbi.nlm.nih.gov/pmc/articles/PMC3818682/} {\bibfield  {journal}
  {\bibinfo  {journal} {Sci. Rep.}\ }\textbf {\bibinfo {volume} {3}},\ \bibinfo
  {pages} {3150} (\bibinfo {year} {2013})}\BibitemShut {NoStop}%
\bibitem [{\citenamefont {Zhu}\ \emph {et~al.}(2013)\citenamefont {Zhu},
  \citenamefont {Nicolaou}, \citenamefont {Levy}, \citenamefont {Butch},
  \citenamefont {Syers}, \citenamefont {Wang}, \citenamefont {Paglione},
  \citenamefont {Sawatzky}, \citenamefont {Elfimov},\ and\ \citenamefont
  {Damascelli}}]{zhu2013polarity}%
  \BibitemOpen
  \bibfield  {author} {\bibinfo {author} {\bibfnamefont {Z.-H.}\ \bibnamefont
  {Zhu}}, \bibinfo {author} {\bibfnamefont {A.}~\bibnamefont {Nicolaou}},
  \bibinfo {author} {\bibfnamefont {G.}~\bibnamefont {Levy}}, \bibinfo {author}
  {\bibfnamefont {N.}~\bibnamefont {Butch}}, \bibinfo {author} {\bibfnamefont
  {P.}~\bibnamefont {Syers}}, \bibinfo {author} {\bibfnamefont
  {X.}~\bibnamefont {Wang}}, \bibinfo {author} {\bibfnamefont {J.}~\bibnamefont
  {Paglione}}, \bibinfo {author} {\bibfnamefont {G.}~\bibnamefont {Sawatzky}},
  \bibinfo {author} {\bibfnamefont {I.}~\bibnamefont {Elfimov}}, \ and\
  \bibinfo {author} {\bibfnamefont {A.}~\bibnamefont {Damascelli}},\ }\href
  {https://journals.aps.org/prl/abstract/10.1103/PhysRevLett.111.216402}
  {\bibfield  {journal} {\bibinfo  {journal} {Phys. Rev. Lett.}\ }\textbf
  {\bibinfo {volume} {111}},\ \bibinfo {pages} {216402} (\bibinfo {year}
  {2013})}\BibitemShut {NoStop}%
\bibitem [{\citenamefont {Wolgast}\ \emph {et~al.}(2015)\citenamefont
  {Wolgast}, \citenamefont {Eo}, \citenamefont {{\"O}zt{\"u}rk}, \citenamefont
  {Li}, \citenamefont {Xiang}, \citenamefont {Tinsman}, \citenamefont {Asaba},
  \citenamefont {Lawson}, \citenamefont {Yu}, \citenamefont {Allen} \emph
  {et~al.}}]{wolgast2015magnetotransport}%
  \BibitemOpen
  \bibfield  {author} {\bibinfo {author} {\bibfnamefont {S.}~\bibnamefont
  {Wolgast}}, \bibinfo {author} {\bibfnamefont {Y.~S.}\ \bibnamefont {Eo}},
  \bibinfo {author} {\bibfnamefont {T.}~\bibnamefont {{\"O}zt{\"u}rk}},
  \bibinfo {author} {\bibfnamefont {G.}~\bibnamefont {Li}}, \bibinfo {author}
  {\bibfnamefont {Z.}~\bibnamefont {Xiang}}, \bibinfo {author} {\bibfnamefont
  {C.}~\bibnamefont {Tinsman}}, \bibinfo {author} {\bibfnamefont
  {T.}~\bibnamefont {Asaba}}, \bibinfo {author} {\bibfnamefont
  {B.}~\bibnamefont {Lawson}}, \bibinfo {author} {\bibfnamefont
  {F.}~\bibnamefont {Yu}}, \bibinfo {author} {\bibfnamefont {J.~W.}\
  \bibnamefont {Allen}},  \emph {et~al.},\ }\href
  {https://journals.aps.org/prb/abstract/10.1103/PhysRevB.92.115110} {\bibfield
   {journal} {\bibinfo  {journal} {Phys. Rev. B}\ }\textbf {\bibinfo {volume}
  {92}},\ \bibinfo {pages} {115110} (\bibinfo {year} {2015})}\BibitemShut
  {NoStop}%
\bibitem [{\citenamefont {Nakajima}\ \emph {et~al.}(2016)\citenamefont
  {Nakajima}, \citenamefont {Syers}, \citenamefont {Wang}, \citenamefont
  {Wang},\ and\ \citenamefont {Paglione}}]{nakajima2016one}%
  \BibitemOpen
  \bibfield  {author} {\bibinfo {author} {\bibfnamefont {Y.}~\bibnamefont
  {Nakajima}}, \bibinfo {author} {\bibfnamefont {P.}~\bibnamefont {Syers}},
  \bibinfo {author} {\bibfnamefont {X.}~\bibnamefont {Wang}}, \bibinfo {author}
  {\bibfnamefont {R.}~\bibnamefont {Wang}}, \ and\ \bibinfo {author}
  {\bibfnamefont {J.}~\bibnamefont {Paglione}},\ }\href
  {https://www.nature.com/articles/nphys3555} {\bibfield  {journal} {\bibinfo
  {journal} {Nat. Phys.}\ }\textbf {\bibinfo {volume} {12}},\ \bibinfo {pages}
  {213} (\bibinfo {year} {2016})}\BibitemShut {NoStop}%
\bibitem [{\citenamefont {Syers}\ \emph {et~al.}(2015)\citenamefont {Syers},
  \citenamefont {Kim}, \citenamefont {Fuhrer},\ and\ \citenamefont
  {Paglione}}]{syers2015tuning}%
  \BibitemOpen
  \bibfield  {author} {\bibinfo {author} {\bibfnamefont {P.}~\bibnamefont
  {Syers}}, \bibinfo {author} {\bibfnamefont {D.}~\bibnamefont {Kim}}, \bibinfo
  {author} {\bibfnamefont {M.~S.}\ \bibnamefont {Fuhrer}}, \ and\ \bibinfo
  {author} {\bibfnamefont {J.}~\bibnamefont {Paglione}},\ }\href
  {https://journals.aps.org/prl/abstract/10.1103/PhysRevLett.114.096601}
  {\bibfield  {journal} {\bibinfo  {journal} {Phys. Rev. Lett.}\ }\textbf
  {\bibinfo {volume} {114}},\ \bibinfo {pages} {096601} (\bibinfo {year}
  {2015})}\BibitemShut {NoStop}%
\bibitem [{\citenamefont {Thomas}\ \emph {et~al.}(2016)\citenamefont {Thomas},
  \citenamefont {Kim}, \citenamefont {Chung}, \citenamefont {Grant},
  \citenamefont {Fisk},\ and\ \citenamefont {Xia}}]{thomas2016weak}%
  \BibitemOpen
  \bibfield  {author} {\bibinfo {author} {\bibfnamefont {S.}~\bibnamefont
  {Thomas}}, \bibinfo {author} {\bibfnamefont {D.}~\bibnamefont {Kim}},
  \bibinfo {author} {\bibfnamefont {S.}~\bibnamefont {Chung}}, \bibinfo
  {author} {\bibfnamefont {T.}~\bibnamefont {Grant}}, \bibinfo {author}
  {\bibfnamefont {Z.}~\bibnamefont {Fisk}}, \ and\ \bibinfo {author}
  {\bibfnamefont {J.}~\bibnamefont {Xia}},\ }\href
  {https://journals.aps.org/prb/abstract/10.1103/PhysRevB.94.205114} {\bibfield
   {journal} {\bibinfo  {journal} {Phys. Rev. B}\ }\textbf {\bibinfo {volume}
  {94}},\ \bibinfo {pages} {205114} (\bibinfo {year} {2016})}\BibitemShut
  {NoStop}%
\bibitem [{\citenamefont {Wakeham}\ \emph {et~al.}(2015)\citenamefont
  {Wakeham}, \citenamefont {Wang}, \citenamefont {Fisk}, \citenamefont
  {Ronning},\ and\ \citenamefont {Thompson}}]{PhysRevB.91.085107}%
  \BibitemOpen
  \bibfield  {author} {\bibinfo {author} {\bibfnamefont {N.}~\bibnamefont
  {Wakeham}}, \bibinfo {author} {\bibfnamefont {Y.~Q.}\ \bibnamefont {Wang}},
  \bibinfo {author} {\bibfnamefont {Z.}~\bibnamefont {Fisk}}, \bibinfo {author}
  {\bibfnamefont {F.}~\bibnamefont {Ronning}}, \ and\ \bibinfo {author}
  {\bibfnamefont {J.~D.}\ \bibnamefont {Thompson}},\ }\href {\doibase
  10.1103/PhysRevB.91.085107} {\bibfield  {journal} {\bibinfo  {journal} {Phys.
  Rev. B}\ }\textbf {\bibinfo {volume} {91}},\ \bibinfo {pages} {085107}
  (\bibinfo {year} {2015})}\BibitemShut {NoStop}%
\bibitem [{\citenamefont {Luo}\ \emph {et~al.}(2015)\citenamefont {Luo},
  \citenamefont {Chen}, \citenamefont {Dai}, \citenamefont {Xu},\ and\
  \citenamefont {Thompson}}]{luo2015heavy}%
  \BibitemOpen
  \bibfield  {author} {\bibinfo {author} {\bibfnamefont {Y.}~\bibnamefont
  {Luo}}, \bibinfo {author} {\bibfnamefont {H.}~\bibnamefont {Chen}}, \bibinfo
  {author} {\bibfnamefont {J.}~\bibnamefont {Dai}}, \bibinfo {author}
  {\bibfnamefont {Z.-a.}\ \bibnamefont {Xu}}, \ and\ \bibinfo {author}
  {\bibfnamefont {J.~D.}\ \bibnamefont {Thompson}},\ }\href
  {https://journals.aps.org/prb/abstract/10.1103/PhysRevB.91.075130} {\bibfield
   {journal} {\bibinfo  {journal} {Phys. Rev. B}\ }\textbf {\bibinfo {volume}
  {91}},\ \bibinfo {pages} {075130} (\bibinfo {year} {2015})}\BibitemShut
  {NoStop}%
\bibitem [{\citenamefont {Li}\ \emph {et~al.}(2014)\citenamefont {Li},
  \citenamefont {Xiang}, \citenamefont {Yu}, \citenamefont {Asaba},
  \citenamefont {Lawson}, \citenamefont {Cai}, \citenamefont {Tinsman},
  \citenamefont {Berkley}, \citenamefont {Wolgast}, \citenamefont {Eo} \emph
  {et~al.}}]{li2014two}%
  \BibitemOpen
  \bibfield  {author} {\bibinfo {author} {\bibfnamefont {G.}~\bibnamefont
  {Li}}, \bibinfo {author} {\bibfnamefont {Z.}~\bibnamefont {Xiang}}, \bibinfo
  {author} {\bibfnamefont {F.}~\bibnamefont {Yu}}, \bibinfo {author}
  {\bibfnamefont {T.}~\bibnamefont {Asaba}}, \bibinfo {author} {\bibfnamefont
  {B.}~\bibnamefont {Lawson}}, \bibinfo {author} {\bibfnamefont
  {P.}~\bibnamefont {Cai}}, \bibinfo {author} {\bibfnamefont {C.}~\bibnamefont
  {Tinsman}}, \bibinfo {author} {\bibfnamefont {A.}~\bibnamefont {Berkley}},
  \bibinfo {author} {\bibfnamefont {S.}~\bibnamefont {Wolgast}}, \bibinfo
  {author} {\bibfnamefont {Y.~S.}\ \bibnamefont {Eo}},  \emph {et~al.},\ }\href
  {http://science.sciencemag.org/content/346/6214/1208} {\bibfield  {journal}
  {\bibinfo  {journal} {Science}\ }\textbf {\bibinfo {volume} {346}},\ \bibinfo
  {pages} {1208} (\bibinfo {year} {2014})}\BibitemShut {NoStop}%
\bibitem [{\citenamefont {Neupane}\ \emph {et~al.}(2013)\citenamefont
  {Neupane}, \citenamefont {Alidoust}, \citenamefont {Xu}, \citenamefont
  {Kondo}, \citenamefont {Ishida}, \citenamefont {Kim}, \citenamefont {Liu},
  \citenamefont {Belopolski}, \citenamefont {Jo}, \citenamefont {Chang} \emph
  {et~al.}}]{neupane2013surface}%
  \BibitemOpen
  \bibfield  {author} {\bibinfo {author} {\bibfnamefont {M.}~\bibnamefont
  {Neupane}}, \bibinfo {author} {\bibfnamefont {N.}~\bibnamefont {Alidoust}},
  \bibinfo {author} {\bibfnamefont {S.}~\bibnamefont {Xu}}, \bibinfo {author}
  {\bibfnamefont {T.}~\bibnamefont {Kondo}}, \bibinfo {author} {\bibfnamefont
  {Y.}~\bibnamefont {Ishida}}, \bibinfo {author} {\bibfnamefont {D.-J.}\
  \bibnamefont {Kim}}, \bibinfo {author} {\bibfnamefont {C.}~\bibnamefont
  {Liu}}, \bibinfo {author} {\bibfnamefont {I.}~\bibnamefont {Belopolski}},
  \bibinfo {author} {\bibfnamefont {Y.}~\bibnamefont {Jo}}, \bibinfo {author}
  {\bibfnamefont {T.-R.}\ \bibnamefont {Chang}},  \emph {et~al.},\ }\href
  {https://www.nature.com/articles/ncomms3991} {\bibfield  {journal} {\bibinfo
  {journal} {Nat. Commun.}\ }\textbf {\bibinfo {volume} {4}},\ \bibinfo {pages}
  {2991} (\bibinfo {year} {2013})}\BibitemShut {NoStop}%
\bibitem [{\citenamefont {Denlinger}\ \emph {et~al.}(2014)\citenamefont
  {Denlinger}, \citenamefont {Allen}, \citenamefont {Kang}, \citenamefont
  {Sun}, \citenamefont {Min}, \citenamefont {Kim},\ and\ \citenamefont
  {Fisk}}]{denlinger2014smb6}%
  \BibitemOpen
  \bibfield  {author} {\bibinfo {author} {\bibfnamefont {J.~D.}\ \bibnamefont
  {Denlinger}}, \bibinfo {author} {\bibfnamefont {J.~W.}\ \bibnamefont
  {Allen}}, \bibinfo {author} {\bibfnamefont {J.-S.}\ \bibnamefont {Kang}},
  \bibinfo {author} {\bibfnamefont {K.}~\bibnamefont {Sun}}, \bibinfo {author}
  {\bibfnamefont {B.-I.}\ \bibnamefont {Min}}, \bibinfo {author} {\bibfnamefont
  {D.-J.}\ \bibnamefont {Kim}}, \ and\ \bibinfo {author} {\bibfnamefont
  {Z.}~\bibnamefont {Fisk}},\ }in\ \href
  {http://journals.jps.jp/doi/abs/10.7566/JPSCP.3.017038} {\emph {\bibinfo
  {booktitle} {JPS Conf. Proc. 3}}}\ (\bibinfo {year} {2014})\ p.\ \bibinfo
  {pages} {017038}\BibitemShut {NoStop}%
\bibitem [{\citenamefont {Jiang}\ \emph {et~al.}(2013)\citenamefont {Jiang},
  \citenamefont {Li}, \citenamefont {Zhang}, \citenamefont {Sun}, \citenamefont
  {Chen}, \citenamefont {Ye}, \citenamefont {Xu}, \citenamefont {Ge},
  \citenamefont {Tan}, \citenamefont {Niu} \emph
  {et~al.}}]{jiang2013observation}%
  \BibitemOpen
  \bibfield  {author} {\bibinfo {author} {\bibfnamefont {J.}~\bibnamefont
  {Jiang}}, \bibinfo {author} {\bibfnamefont {S.}~\bibnamefont {Li}}, \bibinfo
  {author} {\bibfnamefont {T.}~\bibnamefont {Zhang}}, \bibinfo {author}
  {\bibfnamefont {Z.}~\bibnamefont {Sun}}, \bibinfo {author} {\bibfnamefont
  {F.}~\bibnamefont {Chen}}, \bibinfo {author} {\bibfnamefont {Z.}~\bibnamefont
  {Ye}}, \bibinfo {author} {\bibfnamefont {M.}~\bibnamefont {Xu}}, \bibinfo
  {author} {\bibfnamefont {Q.}~\bibnamefont {Ge}}, \bibinfo {author}
  {\bibfnamefont {S.}~\bibnamefont {Tan}}, \bibinfo {author} {\bibfnamefont
  {X.}~\bibnamefont {Niu}},  \emph {et~al.},\ }\href
  {https://www.nature.com/articles/ncomms4010} {\bibfield  {journal} {\bibinfo
  {journal} {Nat. Commun.}\ }\textbf {\bibinfo {volume} {4}},\ \bibinfo {pages}
  {3010} (\bibinfo {year} {2013})}\BibitemShut {NoStop}%
\bibitem [{\citenamefont {Xu}\ \emph {et~al.}(2014{\natexlab{a}})\citenamefont
  {Xu}, \citenamefont {Matt}, \citenamefont {Pomjakushina}, \citenamefont
  {Shi}, \citenamefont {Dhaka}, \citenamefont {Plumb}, \citenamefont
  {Radovi{\'c}}, \citenamefont {Biswas}, \citenamefont {Evtushinsky},
  \citenamefont {Zabolotnyy} \emph {et~al.}}]{NXU1}%
  \BibitemOpen
  \bibfield  {author} {\bibinfo {author} {\bibfnamefont {N.}~\bibnamefont
  {Xu}}, \bibinfo {author} {\bibfnamefont {C.}~\bibnamefont {Matt}}, \bibinfo
  {author} {\bibfnamefont {E.}~\bibnamefont {Pomjakushina}}, \bibinfo {author}
  {\bibfnamefont {X.}~\bibnamefont {Shi}}, \bibinfo {author} {\bibfnamefont
  {R.}~\bibnamefont {Dhaka}}, \bibinfo {author} {\bibfnamefont
  {N.}~\bibnamefont {Plumb}}, \bibinfo {author} {\bibfnamefont
  {M.}~\bibnamefont {Radovi{\'c}}}, \bibinfo {author} {\bibfnamefont
  {P.}~\bibnamefont {Biswas}}, \bibinfo {author} {\bibfnamefont
  {D.}~\bibnamefont {Evtushinsky}}, \bibinfo {author} {\bibfnamefont
  {V.}~\bibnamefont {Zabolotnyy}},  \emph {et~al.},\ }\href
  {https://journals.aps.org/prb/abstract/10.1103/PhysRevB.90.085148} {\bibfield
   {journal} {\bibinfo  {journal} {Phys. Rev. B}\ }\textbf {\bibinfo {volume}
  {90}},\ \bibinfo {pages} {085148} (\bibinfo {year}
  {2014}{\natexlab{a}})}\BibitemShut {NoStop}%
\bibitem [{\citenamefont {Xu}\ \emph {et~al.}(2013)\citenamefont {Xu},
  \citenamefont {Shi}, \citenamefont {Biswas}, \citenamefont {Matt},
  \citenamefont {Dhaka}, \citenamefont {Huang}, \citenamefont {Plumb},
  \citenamefont {Radovi\ifmmode~\acute{c}\else \'{c}\fi{}}, \citenamefont
  {Dil}, \citenamefont {Pomjakushina}, \citenamefont {Conder}, \citenamefont
  {Amato}, \citenamefont {Salman}, \citenamefont {Paul}, \citenamefont {Mesot},
  \citenamefont {Ding},\ and\ \citenamefont {Shi}}]{NXU2}%
  \BibitemOpen
  \bibfield  {author} {\bibinfo {author} {\bibfnamefont {N.}~\bibnamefont
  {Xu}}, \bibinfo {author} {\bibfnamefont {X.}~\bibnamefont {Shi}}, \bibinfo
  {author} {\bibfnamefont {P.~K.}\ \bibnamefont {Biswas}}, \bibinfo {author}
  {\bibfnamefont {C.~E.}\ \bibnamefont {Matt}}, \bibinfo {author}
  {\bibfnamefont {R.~S.}\ \bibnamefont {Dhaka}}, \bibinfo {author}
  {\bibfnamefont {Y.}~\bibnamefont {Huang}}, \bibinfo {author} {\bibfnamefont
  {N.~C.}\ \bibnamefont {Plumb}}, \bibinfo {author} {\bibfnamefont
  {M.}~\bibnamefont {Radovi\ifmmode~\acute{c}\else \'{c}\fi{}}}, \bibinfo
  {author} {\bibfnamefont {J.~H.}\ \bibnamefont {Dil}}, \bibinfo {author}
  {\bibfnamefont {E.}~\bibnamefont {Pomjakushina}}, \bibinfo {author}
  {\bibfnamefont {K.}~\bibnamefont {Conder}}, \bibinfo {author} {\bibfnamefont
  {A.}~\bibnamefont {Amato}}, \bibinfo {author} {\bibfnamefont
  {Z.}~\bibnamefont {Salman}}, \bibinfo {author} {\bibfnamefont {D.~M.}\
  \bibnamefont {Paul}}, \bibinfo {author} {\bibfnamefont {J.}~\bibnamefont
  {Mesot}}, \bibinfo {author} {\bibfnamefont {H.}~\bibnamefont {Ding}}, \ and\
  \bibinfo {author} {\bibfnamefont {M.}~\bibnamefont {Shi}},\ }\href {\doibase
  10.1103/PhysRevB.88.121102} {\bibfield  {journal} {\bibinfo  {journal} {Phys.
  Rev. B}\ }\textbf {\bibinfo {volume} {88}},\ \bibinfo {pages} {121102}
  (\bibinfo {year} {2013})}\BibitemShut {NoStop}%
\bibitem [{\citenamefont {Yee}\ \emph {et~al.}(2013)\citenamefont {Yee},
  \citenamefont {He}, \citenamefont {Soumyanarayanan}, \citenamefont {Kim},
  \citenamefont {Fisk},\ and\ \citenamefont {Hoffman}}]{yee2013imaging}%
  \BibitemOpen
  \bibfield  {author} {\bibinfo {author} {\bibfnamefont {M.~M.}\ \bibnamefont
  {Yee}}, \bibinfo {author} {\bibfnamefont {Y.}~\bibnamefont {He}}, \bibinfo
  {author} {\bibfnamefont {A.}~\bibnamefont {Soumyanarayanan}}, \bibinfo
  {author} {\bibfnamefont {D.-J.}\ \bibnamefont {Kim}}, \bibinfo {author}
  {\bibfnamefont {Z.}~\bibnamefont {Fisk}}, \ and\ \bibinfo {author}
  {\bibfnamefont {J.~E.}\ \bibnamefont {Hoffman}},\ }\href
  {https://arxiv.org/abs/1308.1085} {\bibfield  {journal} {\bibinfo  {journal}
  {arXiv:1308.1085}\ } (\bibinfo {year} {2013})}\BibitemShut {NoStop}%
\bibitem [{\citenamefont {R{\"o}{\ss}ler}\ \emph {et~al.}(2014)\citenamefont
  {R{\"o}{\ss}ler}, \citenamefont {Jang}, \citenamefont {Kim}, \citenamefont
  {Tjeng}, \citenamefont {Fisk}, \citenamefont {Steglich},\ and\ \citenamefont
  {Wirth}}]{rossler2014hybridization}%
  \BibitemOpen
  \bibfield  {author} {\bibinfo {author} {\bibfnamefont {S.}~\bibnamefont
  {R{\"o}{\ss}ler}}, \bibinfo {author} {\bibfnamefont {T.-H.}\ \bibnamefont
  {Jang}}, \bibinfo {author} {\bibfnamefont {D.-J.}\ \bibnamefont {Kim}},
  \bibinfo {author} {\bibfnamefont {L.}~\bibnamefont {Tjeng}}, \bibinfo
  {author} {\bibfnamefont {Z.}~\bibnamefont {Fisk}}, \bibinfo {author}
  {\bibfnamefont {F.}~\bibnamefont {Steglich}}, \ and\ \bibinfo {author}
  {\bibfnamefont {S.}~\bibnamefont {Wirth}},\ }\href
  {http://www.pnas.org/content/111/13/4798.short} {\bibfield  {journal}
  {\bibinfo  {journal} {Proc. Natl. Acad. Sci. U.S.A}\ }\textbf {\bibinfo
  {volume} {111}},\ \bibinfo {pages} {4798} (\bibinfo {year}
  {2014})}\BibitemShut {NoStop}%
\bibitem [{\citenamefont {Miyamachi}\ \emph {et~al.}()\citenamefont
  {Miyamachi}, \citenamefont {Suga}, \citenamefont {Ellguth}, \citenamefont
  {Tusche}, \citenamefont {Schneider}, \citenamefont {Iga},\ and\ \citenamefont
  {Komori}}]{miyamachi2017evidence}%
  \BibitemOpen
  \bibfield  {author} {\bibinfo {author} {\bibfnamefont {T.}~\bibnamefont
  {Miyamachi}}, \bibinfo {author} {\bibfnamefont {S.}~\bibnamefont {Suga}},
  \bibinfo {author} {\bibfnamefont {M.}~\bibnamefont {Ellguth}}, \bibinfo
  {author} {\bibfnamefont {C.}~\bibnamefont {Tusche}}, \bibinfo {author}
  {\bibfnamefont {C.~M.}\ \bibnamefont {Schneider}}, \bibinfo {author}
  {\bibfnamefont {F.}~\bibnamefont {Iga}}, \ and\ \bibinfo {author}
  {\bibfnamefont {F.}~\bibnamefont {Komori}},\ }\href
  {https://www.nature.com/articles/s41598-017-12887-2} {\bibfield  {journal}
  {\bibinfo  {journal} {Sci. Rep.}\ }\textbf {\bibinfo {volume} {7}},\ \bibinfo
  {pages} {12837}}\BibitemShut {NoStop}%
\bibitem [{\citenamefont {Zhang}\ \emph {et~al.}(2013)\citenamefont {Zhang},
  \citenamefont {Butch}, \citenamefont {Syers}, \citenamefont {Ziemak},
  \citenamefont {Greene},\ and\ \citenamefont
  {Paglione}}]{zhang2013hybridization}%
  \BibitemOpen
  \bibfield  {author} {\bibinfo {author} {\bibfnamefont {X.}~\bibnamefont
  {Zhang}}, \bibinfo {author} {\bibfnamefont {N.}~\bibnamefont {Butch}},
  \bibinfo {author} {\bibfnamefont {P.}~\bibnamefont {Syers}}, \bibinfo
  {author} {\bibfnamefont {S.}~\bibnamefont {Ziemak}}, \bibinfo {author}
  {\bibfnamefont {R.~L.}\ \bibnamefont {Greene}}, \ and\ \bibinfo {author}
  {\bibfnamefont {J.}~\bibnamefont {Paglione}},\ }\href
  {https://journals.aps.org/prx/abstract/10.1103/PhysRevX.3.011011} {\bibfield
  {journal} {\bibinfo  {journal} {Phys. Rev. X}\ }\textbf {\bibinfo {volume}
  {3}},\ \bibinfo {pages} {011011} (\bibinfo {year} {2013})}\BibitemShut
  {NoStop}%
\bibitem [{\citenamefont {Park}\ \emph {et~al.}(2016)\citenamefont {Park},
  \citenamefont {Sun}, \citenamefont {Noddings}, \citenamefont {Kim},
  \citenamefont {Fisk},\ and\ \citenamefont {Greene}}]{park2016topological}%
  \BibitemOpen
  \bibfield  {author} {\bibinfo {author} {\bibfnamefont {W.~K.}\ \bibnamefont
  {Park}}, \bibinfo {author} {\bibfnamefont {L.}~\bibnamefont {Sun}}, \bibinfo
  {author} {\bibfnamefont {A.}~\bibnamefont {Noddings}}, \bibinfo {author}
  {\bibfnamefont {D.-J.}\ \bibnamefont {Kim}}, \bibinfo {author} {\bibfnamefont
  {Z.}~\bibnamefont {Fisk}}, \ and\ \bibinfo {author} {\bibfnamefont {L.~H.}\
  \bibnamefont {Greene}},\ }\href
  {https://www.ncbi.nlm.nih.gov/pubmed/27233936} {\bibfield  {journal}
  {\bibinfo  {journal} {Proc. Natl. Acad. Sci. U.S.A.}\ }\textbf {\bibinfo
  {volume} {113}},\ \bibinfo {pages} {6599} (\bibinfo {year}
  {2016})}\BibitemShut {NoStop}%
\bibitem [{\citenamefont {Sun}\ \emph {et~al.}(2017)\citenamefont {Sun},
  \citenamefont {Kim}, \citenamefont {Fisk},\ and\ \citenamefont
  {Park}}]{sun2017planar}%
  \BibitemOpen
  \bibfield  {author} {\bibinfo {author} {\bibfnamefont {L.}~\bibnamefont
  {Sun}}, \bibinfo {author} {\bibfnamefont {D.-J.}\ \bibnamefont {Kim}},
  \bibinfo {author} {\bibfnamefont {Z.}~\bibnamefont {Fisk}}, \ and\ \bibinfo
  {author} {\bibfnamefont {W.}~\bibnamefont {Park}},\ }\href
  {https://journals.aps.org/prb/abstract/10.1103/PhysRevB.95.195129} {\bibfield
   {journal} {\bibinfo  {journal} {Phys. Rev. B}\ }\textbf {\bibinfo {volume}
  {95}},\ \bibinfo {pages} {195129} (\bibinfo {year} {2017})}\BibitemShut
  {NoStop}%
\bibitem [{\citenamefont {Fuhrman}\ \emph {et~al.}(2015)\citenamefont
  {Fuhrman}, \citenamefont {Leiner}, \citenamefont {Nikoli{\'c}}, \citenamefont
  {Granroth}, \citenamefont {Stone}, \citenamefont {Lumsden}, \citenamefont
  {DeBeer-Schmitt}, \citenamefont {Alekseev}, \citenamefont {Mignot},
  \citenamefont {Koohpayeh} \emph {et~al.}}]{fuhrman2015interaction}%
  \BibitemOpen
  \bibfield  {author} {\bibinfo {author} {\bibfnamefont {W.}~\bibnamefont
  {Fuhrman}}, \bibinfo {author} {\bibfnamefont {J.}~\bibnamefont {Leiner}},
  \bibinfo {author} {\bibfnamefont {P.}~\bibnamefont {Nikoli{\'c}}}, \bibinfo
  {author} {\bibfnamefont {G.~E.}\ \bibnamefont {Granroth}}, \bibinfo {author}
  {\bibfnamefont {M.~B.}\ \bibnamefont {Stone}}, \bibinfo {author}
  {\bibfnamefont {M.~D.}\ \bibnamefont {Lumsden}}, \bibinfo {author}
  {\bibfnamefont {L.}~\bibnamefont {DeBeer-Schmitt}}, \bibinfo {author}
  {\bibfnamefont {P.~A.}\ \bibnamefont {Alekseev}}, \bibinfo {author}
  {\bibfnamefont {J.-M.}\ \bibnamefont {Mignot}}, \bibinfo {author}
  {\bibfnamefont {S.}~\bibnamefont {Koohpayeh}},  \emph {et~al.},\ }\href
  {https://journals.aps.org/prl/abstract/10.1103/PhysRevLett.114.036401}
  {\bibfield  {journal} {\bibinfo  {journal} {Phys. Rev. Lett.}\ }\textbf
  {\bibinfo {volume} {114}},\ \bibinfo {pages} {036401} (\bibinfo {year}
  {2015})}\BibitemShut {NoStop}%
\bibitem [{\citenamefont {Xu}\ \emph {et~al.}(2014{\natexlab{b}})\citenamefont
  {Xu}, \citenamefont {Biswas}, \citenamefont {Dhaka}, \citenamefont {Landolt},
  \citenamefont {Muff}, \citenamefont {Matt}, \citenamefont {Shi},
  \citenamefont {Plumb}, \citenamefont {Radovic}, \citenamefont {Pomjakushina}
  \emph {et~al.}}]{xu2014direct}%
  \BibitemOpen
  \bibfield  {author} {\bibinfo {author} {\bibfnamefont {N.}~\bibnamefont
  {Xu}}, \bibinfo {author} {\bibfnamefont {P.}~\bibnamefont {Biswas}}, \bibinfo
  {author} {\bibfnamefont {R.}~\bibnamefont {Dhaka}}, \bibinfo {author}
  {\bibfnamefont {G.}~\bibnamefont {Landolt}}, \bibinfo {author} {\bibfnamefont
  {S.}~\bibnamefont {Muff}}, \bibinfo {author} {\bibfnamefont {C.}~\bibnamefont
  {Matt}}, \bibinfo {author} {\bibfnamefont {X.}~\bibnamefont {Shi}}, \bibinfo
  {author} {\bibfnamefont {N.}~\bibnamefont {Plumb}}, \bibinfo {author}
  {\bibfnamefont {M.}~\bibnamefont {Radovic}}, \bibinfo {author} {\bibfnamefont
  {E.}~\bibnamefont {Pomjakushina}},  \emph {et~al.},\ }\href
  {https://www.nature.com/articles/ncomms5566?origin=ppub} {\bibfield
  {journal} {\bibinfo  {journal} {Nat. Commun.}\ }\textbf {\bibinfo {volume}
  {5}},\ \bibinfo {pages} {4566} (\bibinfo {year}
  {2014}{\natexlab{b}})}\BibitemShut {NoStop}%
\bibitem [{\citenamefont {Song}\ \emph {et~al.}(2016)\citenamefont {Song},
  \citenamefont {Mi}, \citenamefont {Zhao}, \citenamefont {Su}, \citenamefont
  {Yuan}, \citenamefont {Xing}, \citenamefont {Chen}, \citenamefont {Wang},
  \citenamefont {Wu}, \citenamefont {Chen} \emph {et~al.}}]{song2016spin}%
  \BibitemOpen
  \bibfield  {author} {\bibinfo {author} {\bibfnamefont {Q.}~\bibnamefont
  {Song}}, \bibinfo {author} {\bibfnamefont {J.}~\bibnamefont {Mi}}, \bibinfo
  {author} {\bibfnamefont {D.}~\bibnamefont {Zhao}}, \bibinfo {author}
  {\bibfnamefont {T.}~\bibnamefont {Su}}, \bibinfo {author} {\bibfnamefont
  {W.}~\bibnamefont {Yuan}}, \bibinfo {author} {\bibfnamefont {W.}~\bibnamefont
  {Xing}}, \bibinfo {author} {\bibfnamefont {Y.}~\bibnamefont {Chen}}, \bibinfo
  {author} {\bibfnamefont {T.}~\bibnamefont {Wang}}, \bibinfo {author}
  {\bibfnamefont {T.}~\bibnamefont {Wu}}, \bibinfo {author} {\bibfnamefont
  {X.~H.}\ \bibnamefont {Chen}},  \emph {et~al.},\ }\href
  {https://www.nature.com/articles/ncomms13485} {\bibfield  {journal} {\bibinfo
   {journal} {Nat. Commun.}\ }\textbf {\bibinfo {volume} {7}} (\bibinfo {year}
  {2016})}\BibitemShut {NoStop}%
\bibitem [{\citenamefont {Tan}\ \emph {et~al.}(2015)\citenamefont {Tan},
  \citenamefont {Hsu}, \citenamefont {Zeng}, \citenamefont {Hatnean},
  \citenamefont {Harrison}, \citenamefont {Zhu}, \citenamefont {Hartstein},
  \citenamefont {Kiourlappou}, \citenamefont {Srivastava}, \citenamefont
  {Johannes} \emph {et~al.}}]{tan2015unconventional}%
  \BibitemOpen
  \bibfield  {author} {\bibinfo {author} {\bibfnamefont {B.}~\bibnamefont
  {Tan}}, \bibinfo {author} {\bibfnamefont {Y.-T.}\ \bibnamefont {Hsu}},
  \bibinfo {author} {\bibfnamefont {B.}~\bibnamefont {Zeng}}, \bibinfo {author}
  {\bibfnamefont {M.~C.}\ \bibnamefont {Hatnean}}, \bibinfo {author}
  {\bibfnamefont {N.}~\bibnamefont {Harrison}}, \bibinfo {author}
  {\bibfnamefont {Z.}~\bibnamefont {Zhu}}, \bibinfo {author} {\bibfnamefont
  {M.}~\bibnamefont {Hartstein}}, \bibinfo {author} {\bibfnamefont
  {M.}~\bibnamefont {Kiourlappou}}, \bibinfo {author} {\bibfnamefont
  {A.}~\bibnamefont {Srivastava}}, \bibinfo {author} {\bibfnamefont
  {M.}~\bibnamefont {Johannes}},  \emph {et~al.},\ }\href
  {http://science.sciencemag.org/content/349/6245/287} {\bibfield  {journal}
  {\bibinfo  {journal} {Science}\ }\textbf {\bibinfo {volume} {349}},\ \bibinfo
  {pages} {287} (\bibinfo {year} {2015})}\BibitemShut {NoStop}%
\bibitem [{\citenamefont {Erten}\ \emph {et~al.}(2017)\citenamefont {Erten},
  \citenamefont {Chang}, \citenamefont {Coleman},\ and\ \citenamefont
  {Tsvelik}}]{PhysRevLett.119.057603}%
  \BibitemOpen
  \bibfield  {author} {\bibinfo {author} {\bibfnamefont {O.}~\bibnamefont
  {Erten}}, \bibinfo {author} {\bibfnamefont {P.-Y.}\ \bibnamefont {Chang}},
  \bibinfo {author} {\bibfnamefont {P.}~\bibnamefont {Coleman}}, \ and\
  \bibinfo {author} {\bibfnamefont {A.~M.}\ \bibnamefont {Tsvelik}},\ }\href
  {\doibase 10.1103/PhysRevLett.119.057603} {\bibfield  {journal} {\bibinfo
  {journal} {Phys. Rev. Lett.}\ }\textbf {\bibinfo {volume} {119}},\ \bibinfo
  {pages} {057603} (\bibinfo {year} {2017})}\BibitemShut {NoStop}%
\bibitem [{\citenamefont {Knolle}\ and\ \citenamefont
  {Cooper}(2017)}]{PhysRevLett.118.096604}%
  \BibitemOpen
  \bibfield  {author} {\bibinfo {author} {\bibfnamefont {J.}~\bibnamefont
  {Knolle}}\ and\ \bibinfo {author} {\bibfnamefont {N.~R.}\ \bibnamefont
  {Cooper}},\ }\href {\doibase 10.1103/PhysRevLett.118.096604} {\bibfield
  {journal} {\bibinfo  {journal} {Phys. Rev. Lett.}\ }\textbf {\bibinfo
  {volume} {118}},\ \bibinfo {pages} {096604} (\bibinfo {year}
  {2017})}\BibitemShut {NoStop}%
\bibitem [{\citenamefont {Kasuya}\ \emph {et~al.}(1977)\citenamefont {Kasuya},
  \citenamefont {Kojima},\ and\ \citenamefont {Kasaya}}]{kasuya1977theory}%
  \BibitemOpen
  \bibfield  {author} {\bibinfo {author} {\bibfnamefont {T.}~\bibnamefont
  {Kasuya}}, \bibinfo {author} {\bibfnamefont {K.}~\bibnamefont {Kojima}}, \
  and\ \bibinfo {author} {\bibfnamefont {M.}~\bibnamefont {Kasaya}},\ }in\
  \href {https://link.springer.com/chapter/10.1007/978-1-4615-8816-0_13} {\emph
  {\bibinfo {booktitle} {Valence Instabilities and Related Narrow-Band
  Phenomena}}}\ (\bibinfo  {publisher} {Springer},\ \bibinfo {year} {1977})\
  pp.\ \bibinfo {pages} {137--152}\BibitemShut {NoStop}%
\bibitem [{\citenamefont {Eo}\ \emph {et~al.}(2017)\citenamefont {Eo},
  \citenamefont {Sun}, \citenamefont {Kurdak}, \citenamefont {Kim},\ and\
  \citenamefont {Fisk}}]{eo2017new}%
  \BibitemOpen
  \bibfield  {author} {\bibinfo {author} {\bibfnamefont {Y.}~\bibnamefont
  {Eo}}, \bibinfo {author} {\bibfnamefont {K.}~\bibnamefont {Sun}}, \bibinfo
  {author} {\bibfnamefont {{\c{C}}.}~\bibnamefont {Kurdak}}, \bibinfo {author}
  {\bibfnamefont {D.-J.}\ \bibnamefont {Kim}}, \ and\ \bibinfo {author}
  {\bibfnamefont {Z.}~\bibnamefont {Fisk}},\ }\href
  {https://arxiv.org/abs/1708.05762} {\bibfield  {journal} {\bibinfo  {journal}
  {arXiv:1708.05762}\ } (\bibinfo {year} {2017})}\BibitemShut {NoStop}%
\bibitem [{\citenamefont {Molnar}\ \emph {et~al.}(1982)\citenamefont {Molnar},
  \citenamefont {Theis}, \citenamefont {Benoit}, \citenamefont {Briggs},
  \citenamefont {Flouquet}, \citenamefont {Ravex},\ and\ \citenamefont
  {Fisk}}]{molnar1982study}%
  \BibitemOpen
  \bibfield  {author} {\bibinfo {author} {\bibfnamefont {S.~v.}\ \bibnamefont
  {Molnar}}, \bibinfo {author} {\bibfnamefont {T.}~\bibnamefont {Theis}},
  \bibinfo {author} {\bibfnamefont {A.}~\bibnamefont {Benoit}}, \bibinfo
  {author} {\bibfnamefont {A.}~\bibnamefont {Briggs}}, \bibinfo {author}
  {\bibfnamefont {J.}~\bibnamefont {Flouquet}}, \bibinfo {author}
  {\bibfnamefont {J.}~\bibnamefont {Ravex}}, \ and\ \bibinfo {author}
  {\bibfnamefont {Z.}~\bibnamefont {Fisk}},\ }in\ \href
  {https://inis.iaea.org/search/search.aspx?orig_q=RN:14797406} {\emph
  {\bibinfo {booktitle} {Valence instabilities}}}\ (\bibinfo {year}
  {1982})\BibitemShut {NoStop}%
\bibitem [{\citenamefont {Kasuya}\ \emph {et~al.}(1983)\citenamefont {Kasuya},
  \citenamefont {Kasaya}, \citenamefont {Takegahara}, \citenamefont {Fujita},
  \citenamefont {Goto}, \citenamefont {Tamaki}, \citenamefont {Takigawa},\ and\
  \citenamefont {Yasuoka}}]{kasuya1983mechanisms}%
  \BibitemOpen
  \bibfield  {author} {\bibinfo {author} {\bibfnamefont {T.}~\bibnamefont
  {Kasuya}}, \bibinfo {author} {\bibfnamefont {M.}~\bibnamefont {Kasaya}},
  \bibinfo {author} {\bibfnamefont {K.}~\bibnamefont {Takegahara}}, \bibinfo
  {author} {\bibfnamefont {T.}~\bibnamefont {Fujita}}, \bibinfo {author}
  {\bibfnamefont {T.}~\bibnamefont {Goto}}, \bibinfo {author} {\bibfnamefont
  {A.}~\bibnamefont {Tamaki}}, \bibinfo {author} {\bibfnamefont
  {M.}~\bibnamefont {Takigawa}}, \ and\ \bibinfo {author} {\bibfnamefont
  {H.}~\bibnamefont {Yasuoka}},\ }\href
  {https://www.sciencedirect.com/science/article/pii/0304885383903153}
  {\bibfield  {journal} {\bibinfo  {journal} {J. Magn. Magn. Mater}\ }\textbf
  {\bibinfo {volume} {31}},\ \bibinfo {pages} {447} (\bibinfo {year}
  {1983})}\BibitemShut {NoStop}%
\bibitem [{\citenamefont {Sluchanko}\ \emph {et~al.}(1999)\citenamefont
  {Sluchanko}, \citenamefont {Volkov}, \citenamefont {Glushkov}, \citenamefont
  {Gorshunov}, \citenamefont {Demishev}, \citenamefont {Kondrin}, \citenamefont
  {Pronin}, \citenamefont {Samarin}, \citenamefont {Bruynseraede},
  \citenamefont {Moshchalkov} \emph {et~al.}}]{sluchanko1999nature}%
  \BibitemOpen
  \bibfield  {author} {\bibinfo {author} {\bibfnamefont {N.}~\bibnamefont
  {Sluchanko}}, \bibinfo {author} {\bibfnamefont {A.}~\bibnamefont {Volkov}},
  \bibinfo {author} {\bibfnamefont {V.}~\bibnamefont {Glushkov}}, \bibinfo
  {author} {\bibfnamefont {B.}~\bibnamefont {Gorshunov}}, \bibinfo {author}
  {\bibfnamefont {S.}~\bibnamefont {Demishev}}, \bibinfo {author}
  {\bibfnamefont {M.}~\bibnamefont {Kondrin}}, \bibinfo {author} {\bibfnamefont
  {A.}~\bibnamefont {Pronin}}, \bibinfo {author} {\bibfnamefont
  {N.}~\bibnamefont {Samarin}}, \bibinfo {author} {\bibfnamefont
  {Y.}~\bibnamefont {Bruynseraede}}, \bibinfo {author} {\bibfnamefont
  {V.}~\bibnamefont {Moshchalkov}},  \emph {et~al.},\ }\href
  {https://link.springer.com/article/10.1134/1.558825} {\bibfield  {journal}
  {\bibinfo  {journal} {J. Exp. Theor. Phys}\ }\textbf {\bibinfo {volume}
  {88}},\ \bibinfo {pages} {533} (\bibinfo {year} {1999})}\BibitemShut
  {NoStop}%
\bibitem [{\citenamefont {Sluchanko}\ \emph {et~al.}(2000)\citenamefont
  {Sluchanko}, \citenamefont {\ifmmode~\dot{G}\else \.{G}\fi{}lushkov},
  \citenamefont {Gorshunov}, \citenamefont {Demishev}, \citenamefont {Kondrin},
  \citenamefont {Pronin}, \citenamefont {Volkov}, \citenamefont {Savchenko},
  \citenamefont {Gr\"uner}, \citenamefont {Bruynseraede}, \citenamefont
  {Moshchalkov},\ and\ \citenamefont {Kunii}}]{PhysRevB.61.9906}%
  \BibitemOpen
  \bibfield  {author} {\bibinfo {author} {\bibfnamefont {N.~E.}\ \bibnamefont
  {Sluchanko}}, \bibinfo {author} {\bibfnamefont {V.~V.}\ \bibnamefont
  {\ifmmode~\dot{G}\else \.{G}\fi{}lushkov}}, \bibinfo {author} {\bibfnamefont
  {B.~P.}\ \bibnamefont {Gorshunov}}, \bibinfo {author} {\bibfnamefont {S.~V.}\
  \bibnamefont {Demishev}}, \bibinfo {author} {\bibfnamefont {M.~V.}\
  \bibnamefont {Kondrin}}, \bibinfo {author} {\bibfnamefont {A.~A.}\
  \bibnamefont {Pronin}}, \bibinfo {author} {\bibfnamefont {A.~A.}\
  \bibnamefont {Volkov}}, \bibinfo {author} {\bibfnamefont {A.~K.}\
  \bibnamefont {Savchenko}}, \bibinfo {author} {\bibfnamefont {G.}~\bibnamefont
  {Gr\"uner}}, \bibinfo {author} {\bibfnamefont {Y.}~\bibnamefont
  {Bruynseraede}}, \bibinfo {author} {\bibfnamefont {V.~V.}\ \bibnamefont
  {Moshchalkov}}, \ and\ \bibinfo {author} {\bibfnamefont {S.}~\bibnamefont
  {Kunii}},\ }\href {\doibase 10.1103/PhysRevB.61.9906} {\bibfield  {journal}
  {\bibinfo  {journal} {Phys. Rev. B}\ }\textbf {\bibinfo {volume} {61}},\
  \bibinfo {pages} {9906} (\bibinfo {year} {2000})}\BibitemShut {NoStop}%
\bibitem [{\citenamefont {Laurita}\ \emph {et~al.}(2016)\citenamefont
  {Laurita}, \citenamefont {Morris}, \citenamefont {Koohpayeh}, \citenamefont
  {Rosa}, \citenamefont {Phelan}, \citenamefont {Fisk}, \citenamefont
  {McQueen},\ and\ \citenamefont {Armitage}}]{PhysRevB.94.165154}%
  \BibitemOpen
  \bibfield  {author} {\bibinfo {author} {\bibfnamefont {N.~J.}\ \bibnamefont
  {Laurita}}, \bibinfo {author} {\bibfnamefont {C.~M.}\ \bibnamefont {Morris}},
  \bibinfo {author} {\bibfnamefont {S.~M.}\ \bibnamefont {Koohpayeh}}, \bibinfo
  {author} {\bibfnamefont {P.~F.~S.}\ \bibnamefont {Rosa}}, \bibinfo {author}
  {\bibfnamefont {W.~A.}\ \bibnamefont {Phelan}}, \bibinfo {author}
  {\bibfnamefont {Z.}~\bibnamefont {Fisk}}, \bibinfo {author} {\bibfnamefont
  {T.~M.}\ \bibnamefont {McQueen}}, \ and\ \bibinfo {author} {\bibfnamefont
  {N.~P.}\ \bibnamefont {Armitage}},\ }\href {\doibase
  10.1103/PhysRevB.94.165154} {\bibfield  {journal} {\bibinfo  {journal} {Phys.
  Rev. B}\ }\textbf {\bibinfo {volume} {94}},\ \bibinfo {pages} {165154}
  (\bibinfo {year} {2016})}\BibitemShut {NoStop}%
\bibitem [{\citenamefont {Laurita}\ \emph {et~al.}(2017)\citenamefont
  {Laurita}, \citenamefont {Morris}, \citenamefont {Koohpayeh}, \citenamefont
  {Phelan}, \citenamefont {McQueen},\ and\ \citenamefont
  {Armitage}}]{laurita2017impurities}%
  \BibitemOpen
  \bibfield  {author} {\bibinfo {author} {\bibfnamefont {N.}~\bibnamefont
  {Laurita}}, \bibinfo {author} {\bibfnamefont {C.}~\bibnamefont {Morris}},
  \bibinfo {author} {\bibfnamefont {S.}~\bibnamefont {Koohpayeh}}, \bibinfo
  {author} {\bibfnamefont {W.}~\bibnamefont {Phelan}}, \bibinfo {author}
  {\bibfnamefont {T.}~\bibnamefont {McQueen}}, \ and\ \bibinfo {author}
  {\bibfnamefont {N.}~\bibnamefont {Armitage}},\ }\href
  {https://www.sciencedirect.com/science/article/pii/S0921452617306026}
  {\bibfield  {journal} {\bibinfo  {journal} {Physica B Condens Matter}\ }
  (\bibinfo {year} {2017})}\BibitemShut {NoStop}%
\bibitem [{\citenamefont {Stankiewicz}\ \emph {et~al.}(2016)\citenamefont
  {Stankiewicz}, \citenamefont {Rosa}, \citenamefont {Schlottmann},\ and\
  \citenamefont {Fisk}}]{stankiewicz2016electrical}%
  \BibitemOpen
  \bibfield  {author} {\bibinfo {author} {\bibfnamefont {J.}~\bibnamefont
  {Stankiewicz}}, \bibinfo {author} {\bibfnamefont {P.~F.}\ \bibnamefont
  {Rosa}}, \bibinfo {author} {\bibfnamefont {P.}~\bibnamefont {Schlottmann}}, \
  and\ \bibinfo {author} {\bibfnamefont {Z.}~\bibnamefont {Fisk}},\ }\href
  {https://journals.aps.org/prb/abstract/10.1103/PhysRevB.94.125141} {\bibfield
   {journal} {\bibinfo  {journal} {Phys. Rev. B}\ }\textbf {\bibinfo {volume}
  {94}},\ \bibinfo {pages} {125141} (\bibinfo {year} {2016})}\BibitemShut
  {NoStop}%
\bibitem [{\citenamefont {Rhyee}\ and\ \citenamefont
  {Cho}(2004)}]{rhyee2004effect}%
  \BibitemOpen
  \bibfield  {author} {\bibinfo {author} {\bibfnamefont {J.-S.}\ \bibnamefont
  {Rhyee}}\ and\ \bibinfo {author} {\bibfnamefont {B.-K.}\ \bibnamefont
  {Cho}},\ }\href {http://aip.scitation.org/doi/abs/10.1063/1.1667834}
  {\bibfield  {journal} {\bibinfo  {journal} {J. Appl. Phys}\ }\textbf
  {\bibinfo {volume} {95}},\ \bibinfo {pages} {6675} (\bibinfo {year}
  {2004})}\BibitemShut {NoStop}%
\bibitem [{\citenamefont {Vonlanthen}\ \emph {et~al.}(2000)\citenamefont
  {Vonlanthen}, \citenamefont {Felder}, \citenamefont {Degiorgi}, \citenamefont
  {Ott}, \citenamefont {Young}, \citenamefont {Bianchi},\ and\ \citenamefont
  {Fisk}}]{PhysRevB.62.10076}%
  \BibitemOpen
  \bibfield  {author} {\bibinfo {author} {\bibfnamefont {P.}~\bibnamefont
  {Vonlanthen}}, \bibinfo {author} {\bibfnamefont {E.}~\bibnamefont {Felder}},
  \bibinfo {author} {\bibfnamefont {L.}~\bibnamefont {Degiorgi}}, \bibinfo
  {author} {\bibfnamefont {H.~R.}\ \bibnamefont {Ott}}, \bibinfo {author}
  {\bibfnamefont {D.~P.}\ \bibnamefont {Young}}, \bibinfo {author}
  {\bibfnamefont {A.~D.}\ \bibnamefont {Bianchi}}, \ and\ \bibinfo {author}
  {\bibfnamefont {Z.}~\bibnamefont {Fisk}},\ }\href {\doibase
  10.1103/PhysRevB.62.10076} {\bibfield  {journal} {\bibinfo  {journal} {Phys.
  Rev. B}\ }\textbf {\bibinfo {volume} {62}},\ \bibinfo {pages} {10076}
  (\bibinfo {year} {2000})}\BibitemShut {NoStop}%
\bibitem [{\citenamefont {Ott}\ \emph {et~al.}(1997)\citenamefont {Ott},
  \citenamefont {Chernikov}, \citenamefont {Felder}, \citenamefont {Degiorgi},
  \citenamefont {Moshopoulou}, \citenamefont {Sarrao},\ and\ \citenamefont
  {Fisk}}]{ott1997structure}%
  \BibitemOpen
  \bibfield  {author} {\bibinfo {author} {\bibfnamefont {H.}~\bibnamefont
  {Ott}}, \bibinfo {author} {\bibfnamefont {M.}~\bibnamefont {Chernikov}},
  \bibinfo {author} {\bibfnamefont {E.}~\bibnamefont {Felder}}, \bibinfo
  {author} {\bibfnamefont {L.}~\bibnamefont {Degiorgi}}, \bibinfo {author}
  {\bibfnamefont {E.}~\bibnamefont {Moshopoulou}}, \bibinfo {author}
  {\bibfnamefont {J.}~\bibnamefont {Sarrao}}, \ and\ \bibinfo {author}
  {\bibfnamefont {Z.}~\bibnamefont {Fisk}},\ }\href
  {https://link.springer.com/article/10.1007/s002570050297} {\bibfield
  {journal} {\bibinfo  {journal} {Zeitschrift f{\"u}r Physik B Condensed
  Matter}\ }\textbf {\bibinfo {volume} {102}},\ \bibinfo {pages} {337}
  (\bibinfo {year} {1997})}\BibitemShut {NoStop}%
\bibitem [{\citenamefont {Kim}\ \emph {et~al.}(2007)\citenamefont {Kim},
  \citenamefont {Sung},\ and\ \citenamefont {Cho}}]{kim2007weak}%
  \BibitemOpen
  \bibfield  {author} {\bibinfo {author} {\bibfnamefont {J.}~\bibnamefont
  {Kim}}, \bibinfo {author} {\bibfnamefont {N.}~\bibnamefont {Sung}}, \ and\
  \bibinfo {author} {\bibfnamefont {B.}~\bibnamefont {Cho}},\ }\href@noop {}
  {\bibfield  {journal} {\bibinfo  {journal} {J. Appl. Phys.}\ }\textbf
  {\bibinfo {volume} {101}},\ \bibinfo {pages} {09D512} (\bibinfo {year}
  {2007})}\BibitemShut {NoStop}%
\bibitem [{\citenamefont {Tarascon}\ \emph {et~al.}(1980)\citenamefont
  {Tarascon}, \citenamefont {Etourneau}, \citenamefont {Dordor}, \citenamefont
  {Hagenmuller}, \citenamefont {Kasaya},\ and\ \citenamefont
  {Coey}}]{tarascon1980magnetic}%
  \BibitemOpen
  \bibfield  {author} {\bibinfo {author} {\bibfnamefont {J.-M.}\ \bibnamefont
  {Tarascon}}, \bibinfo {author} {\bibfnamefont {J.}~\bibnamefont {Etourneau}},
  \bibinfo {author} {\bibfnamefont {P.}~\bibnamefont {Dordor}}, \bibinfo
  {author} {\bibfnamefont {P.}~\bibnamefont {Hagenmuller}}, \bibinfo {author}
  {\bibfnamefont {M.}~\bibnamefont {Kasaya}}, \ and\ \bibinfo {author}
  {\bibfnamefont {J.}~\bibnamefont {Coey}},\ }\href
  {http://aip.scitation.org/doi/abs/10.1063/1.327309} {\bibfield  {journal}
  {\bibinfo  {journal} {J. Appl. Phys}\ }\textbf {\bibinfo {volume} {51}},\
  \bibinfo {pages} {574} (\bibinfo {year} {1980})}\BibitemShut {NoStop}%
\bibitem [{\citenamefont {Wigger}\ \emph {et~al.}(2004)\citenamefont {Wigger},
  \citenamefont {Monnier}, \citenamefont {Ott}, \citenamefont {Young},\ and\
  \citenamefont {Fisk}}]{wigger2004electronic}%
  \BibitemOpen
  \bibfield  {author} {\bibinfo {author} {\bibfnamefont {G.}~\bibnamefont
  {Wigger}}, \bibinfo {author} {\bibfnamefont {R.}~\bibnamefont {Monnier}},
  \bibinfo {author} {\bibfnamefont {H.}~\bibnamefont {Ott}}, \bibinfo {author}
  {\bibfnamefont {D.}~\bibnamefont {Young}}, \ and\ \bibinfo {author}
  {\bibfnamefont {Z.}~\bibnamefont {Fisk}},\ }\href
  {https://journals.aps.org/prb/abstract/10.1103/PhysRevB.69.125118} {\bibfield
   {journal} {\bibinfo  {journal} {Phys. Rev. B}\ }\textbf {\bibinfo {volume}
  {69}},\ \bibinfo {pages} {125118} (\bibinfo {year} {2004})}\BibitemShut
  {NoStop}%
\bibitem [{\citenamefont {Kasuya}\ \emph {et~al.}(1980)\citenamefont {Kasuya},
  \citenamefont {Takegahara}, \citenamefont {Kasaya}, \citenamefont {Isikawa},\
  and\ \citenamefont {Fujita}}]{kasuya1980iv}%
  \BibitemOpen
  \bibfield  {author} {\bibinfo {author} {\bibfnamefont {T.}~\bibnamefont
  {Kasuya}}, \bibinfo {author} {\bibfnamefont {K.}~\bibnamefont {Takegahara}},
  \bibinfo {author} {\bibfnamefont {M.}~\bibnamefont {Kasaya}}, \bibinfo
  {author} {\bibfnamefont {Y.}~\bibnamefont {Isikawa}}, \ and\ \bibinfo
  {author} {\bibfnamefont {T.}~\bibnamefont {Fujita}},\ }\href
  {https://jphyscol.journaldephysique.org/articles/jphyscol/abs/1980/05/jphyscol198041C528/jphyscol198041C528.html}
  {\bibfield  {journal} {\bibinfo  {journal} {Le Journal de Physique
  Colloques}\ }\textbf {\bibinfo {volume} {41}},\ \bibinfo {pages} {C5}
  (\bibinfo {year} {1980})}\BibitemShut {NoStop}%
\bibitem [{\citenamefont {Ando}(2013)}]{ando2013topological}%
  \BibitemOpen
  \bibfield  {author} {\bibinfo {author} {\bibfnamefont {Y.}~\bibnamefont
  {Ando}},\ }\href {http://journals.jps.jp/doi/abs/10.7566/JPSJ.82.102001}
  {\bibfield  {journal} {\bibinfo  {journal} {J. Phys. Soc. Jpn}\ }\textbf
  {\bibinfo {volume} {82}},\ \bibinfo {pages} {102001} (\bibinfo {year}
  {2013})}\BibitemShut {NoStop}%
\bibitem [{\citenamefont {Rakoski}\ \emph {et~al.}(2017)\citenamefont
  {Rakoski}, \citenamefont {Eo}, \citenamefont {Sun},\ and\ \citenamefont
  {Kurdak}}]{PhysRevB.95.195133}%
  \BibitemOpen
  \bibfield  {author} {\bibinfo {author} {\bibfnamefont {A.}~\bibnamefont
  {Rakoski}}, \bibinfo {author} {\bibfnamefont {Y.~S.}\ \bibnamefont {Eo}},
  \bibinfo {author} {\bibfnamefont {K.}~\bibnamefont {Sun}}, \ and\ \bibinfo
  {author} {\bibfnamefont {i.~m.~c.}\ \bibnamefont {Kurdak}},\ }\href {\doibase
  10.1103/PhysRevB.95.195133} {\bibfield  {journal} {\bibinfo  {journal} {Phys.
  Rev. B}\ }\textbf {\bibinfo {volume} {95}},\ \bibinfo {pages} {195133}
  (\bibinfo {year} {2017})}\BibitemShut {NoStop}%
\bibitem [{\citenamefont {Anderson}(1959)}]{anderson1959theory}%
  \BibitemOpen
  \bibfield  {author} {\bibinfo {author} {\bibfnamefont {P.~W.}\ \bibnamefont
  {Anderson}},\ }\href
  {http://www.sciencedirect.com/science/article/pii/0022369759900368}
  {\bibfield  {journal} {\bibinfo  {journal} {J. Phys. Chem. Solids}\ }\textbf
  {\bibinfo {volume} {11}} (\bibinfo {year} {1959})}\BibitemShut {NoStop}%
\bibitem [{\citenamefont {Ran}\ \emph {et~al.}(2009)\citenamefont {Ran},
  \citenamefont {Zhang},\ and\ \citenamefont {Vishwanath}}]{ran2009one}%
  \BibitemOpen
  \bibfield  {author} {\bibinfo {author} {\bibfnamefont {Y.}~\bibnamefont
  {Ran}}, \bibinfo {author} {\bibfnamefont {Y.}~\bibnamefont {Zhang}}, \ and\
  \bibinfo {author} {\bibfnamefont {A.}~\bibnamefont {Vishwanath}},\ }\href
  {https://www.nature.com/articles/nphys1220} {\bibfield  {journal} {\bibinfo
  {journal} {Nat. Phys.}\ }\textbf {\bibinfo {volume} {5}} (\bibinfo {year}
  {2009})}\BibitemShut {NoStop}%
\bibitem [{\citenamefont {Valentine}\ \emph {et~al.}(2017)\citenamefont
  {Valentine}, \citenamefont {Koohpayeh}, \citenamefont {Phelan}, \citenamefont
  {McQueen}, \citenamefont {Rosa}, \citenamefont {Fisk},\ and\ \citenamefont
  {Drichko}}]{valentine2017effect}%
  \BibitemOpen
  \bibfield  {author} {\bibinfo {author} {\bibfnamefont {M.~E.}\ \bibnamefont
  {Valentine}}, \bibinfo {author} {\bibfnamefont {S.}~\bibnamefont
  {Koohpayeh}}, \bibinfo {author} {\bibfnamefont {W.~A.}\ \bibnamefont
  {Phelan}}, \bibinfo {author} {\bibfnamefont {T.~M.}\ \bibnamefont {McQueen}},
  \bibinfo {author} {\bibfnamefont {P.~F.}\ \bibnamefont {Rosa}}, \bibinfo
  {author} {\bibfnamefont {Z.}~\bibnamefont {Fisk}}, \ and\ \bibinfo {author}
  {\bibfnamefont {N.}~\bibnamefont {Drichko}},\ }\href
  {https://www.sciencedirect.com/science/article/pii/S0921452617307846}
  {\bibfield  {journal} {\bibinfo  {journal} {Physica B Condens Matter}\ }
  (\bibinfo {year} {2017})}\BibitemShut {NoStop}%
\bibitem [{\citenamefont {Fuhrman}\ \emph {et~al.}(2017)\citenamefont
  {Fuhrman}, \citenamefont {Chamorro}, \citenamefont {Alekseev}, \citenamefont
  {Mignot}, \citenamefont {Keller}, \citenamefont {Nikolic}, \citenamefont
  {McQueen},\ and\ \citenamefont {Broholm}}]{fuhrman2017screened}%
  \BibitemOpen
  \bibfield  {author} {\bibinfo {author} {\bibfnamefont {W.~T.}\ \bibnamefont
  {Fuhrman}}, \bibinfo {author} {\bibfnamefont {J.~R.}\ \bibnamefont
  {Chamorro}}, \bibinfo {author} {\bibfnamefont {P.~A.}\ \bibnamefont
  {Alekseev}}, \bibinfo {author} {\bibfnamefont {J.-M.}\ \bibnamefont
  {Mignot}}, \bibinfo {author} {\bibfnamefont {T.}~\bibnamefont {Keller}},
  \bibinfo {author} {\bibfnamefont {P.}~\bibnamefont {Nikolic}}, \bibinfo
  {author} {\bibfnamefont {T.~M.}\ \bibnamefont {McQueen}}, \ and\ \bibinfo
  {author} {\bibfnamefont {C.~L.}\ \bibnamefont {Broholm}},\ }\href
  {https://arxiv.org/abs/1707.03834} {\bibfield  {journal} {\bibinfo  {journal}
  {arXiv:1707.03834}\ } (\bibinfo {year} {2017})}\BibitemShut {NoStop}%
\bibitem [{\citenamefont {Boulanger}\ \emph {et~al.}(2017)\citenamefont
  {Boulanger}, \citenamefont {Lalibert{\'e}}, \citenamefont {Badoux},
  \citenamefont {Doiron-Leyraud}, \citenamefont {Phelan}, \citenamefont
  {Koohpayeh}, \citenamefont {McQueen}, \citenamefont {Wang}, \citenamefont
  {Nakajima}, \citenamefont {Metz} \emph {et~al.}}]{boulanger2017field}%
  \BibitemOpen
  \bibfield  {author} {\bibinfo {author} {\bibfnamefont {M.}~\bibnamefont
  {Boulanger}}, \bibinfo {author} {\bibfnamefont {F.}~\bibnamefont
  {Lalibert{\'e}}}, \bibinfo {author} {\bibfnamefont {S.}~\bibnamefont
  {Badoux}}, \bibinfo {author} {\bibfnamefont {N.}~\bibnamefont
  {Doiron-Leyraud}}, \bibinfo {author} {\bibfnamefont {W.}~\bibnamefont
  {Phelan}}, \bibinfo {author} {\bibfnamefont {S.}~\bibnamefont {Koohpayeh}},
  \bibinfo {author} {\bibfnamefont {T.}~\bibnamefont {McQueen}}, \bibinfo
  {author} {\bibfnamefont {X.}~\bibnamefont {Wang}}, \bibinfo {author}
  {\bibfnamefont {Y.}~\bibnamefont {Nakajima}}, \bibinfo {author}
  {\bibfnamefont {T.}~\bibnamefont {Metz}},  \emph {et~al.},\ }\href
  {https://arxiv.org/abs/1709.10456} {\bibfield  {journal} {\bibinfo  {journal}
  {arXiv:1709.10456}\ } (\bibinfo {year} {2017})}\BibitemShut {NoStop}%
\end{thebibliography}%


\begin{thebibliography}{6}%
\makeatletter
\providecommand \@ifxundefined [1]{%
 \@ifx{#1\undefined}
}%
\providecommand \@ifnum [1]{%
 \ifnum #1\expandafter \@firstoftwo
 \else \expandafter \@secondoftwo
 \fi
}%
\providecommand \@ifx [1]{%
 \ifx #1\expandafter \@firstoftwo
 \else \expandafter \@secondoftwo
 \fi
}%
\providecommand \natexlab [1]{#1}%
\providecommand \enquote  [1]{``#1''}%
\providecommand \bibnamefont  [1]{#1}%
\providecommand \bibfnamefont [1]{#1}%
\providecommand \citenamefont [1]{#1}%
\providecommand \href@noop [0]{\@secondoftwo}%
\providecommand \href [0]{\begingroup \@sanitize@url \@href}%
\providecommand \@href[1]{\@@startlink{#1}\@@href}%
\providecommand \@@href[1]{\endgroup#1\@@endlink}%
\providecommand \@sanitize@url [0]{\catcode `\\12\catcode `\$12\catcode
  `\&12\catcode `\#12\catcode `\^12\catcode `\_12\catcode `\%12\relax}%
\providecommand \@@startlink[1]{}%
\providecommand \@@endlink[0]{}%
\providecommand \url  [0]{\begingroup\@sanitize@url \@url }%
\providecommand \@url [1]{\endgroup\@href {#1}{\urlprefix }}%
\providecommand \urlprefix  [0]{URL }%
\providecommand \Eprint [0]{\href }%
\providecommand \doibase [0]{http://dx.doi.org/}%
\providecommand \selectlanguage [0]{\@gobble}%
\providecommand \bibinfo  [0]{\@secondoftwo}%
\providecommand \bibfield  [0]{\@secondoftwo}%
\providecommand \translation [1]{[#1]}%
\providecommand \BibitemOpen [0]{}%
\providecommand \bibitemStop [0]{}%
\providecommand \bibitemNoStop [0]{.\EOS\space}%
\providecommand \EOS [0]{\spacefactor3000\relax}%
\providecommand \BibitemShut  [1]{\csname bibitem#1\endcsname}%
\let\auto@bib@innerbib\@empty
\bibitem [{\citenamefont {Wolgast}\ \emph
  {et~al.}(2015{\natexlab{a}})\citenamefont {Wolgast}, \citenamefont {Eo},
  \citenamefont {{\"O}zt{\"u}rk}, \citenamefont {Li}, \citenamefont {Xiang},
  \citenamefont {Tinsman}, \citenamefont {Asaba}, \citenamefont {Lawson},
  \citenamefont {Yu}, \citenamefont {Allen} \emph
  {et~al.}}]{wolgast2015magnetotransport}%
  \BibitemOpen
  \bibfield  {author} {\bibinfo {author} {\bibfnamefont {S.}~\bibnamefont
  {Wolgast}}, \bibinfo {author} {\bibfnamefont {Y.~S.}\ \bibnamefont {Eo}},
  \bibinfo {author} {\bibfnamefont {T.}~\bibnamefont {{\"O}zt{\"u}rk}},
  \bibinfo {author} {\bibfnamefont {G.}~\bibnamefont {Li}}, \bibinfo {author}
  {\bibfnamefont {Z.}~\bibnamefont {Xiang}}, \bibinfo {author} {\bibfnamefont
  {C.}~\bibnamefont {Tinsman}}, \bibinfo {author} {\bibfnamefont
  {T.}~\bibnamefont {Asaba}}, \bibinfo {author} {\bibfnamefont
  {B.}~\bibnamefont {Lawson}}, \bibinfo {author} {\bibfnamefont
  {F.}~\bibnamefont {Yu}}, \bibinfo {author} {\bibfnamefont {J.~W.}\
  \bibnamefont {Allen}},  \emph {et~al.},\ }\href
  {https://journals.aps.org/prb/abstract/10.1103/PhysRevB.92.115110} {\bibfield
   {journal} {\bibinfo  {journal} {Phys. Rev. B}\ }\textbf {\bibinfo {volume}
  {92}},\ \bibinfo {pages} {115110} (\bibinfo {year}
  {2015}{\natexlab{a}})}\BibitemShut {NoStop}%
\bibitem [{\citenamefont {Wolgast}\ \emph
  {et~al.}(2015{\natexlab{b}})\citenamefont {Wolgast}, \citenamefont {Eo},
  \citenamefont {Kurdak}, \citenamefont {Kim},\ and\ \citenamefont
  {Fisk}}]{wolgast2015conduction}%
  \BibitemOpen
  \bibfield  {author} {\bibinfo {author} {\bibfnamefont {S.}~\bibnamefont
  {Wolgast}}, \bibinfo {author} {\bibfnamefont {Y.}~\bibnamefont {Eo}},
  \bibinfo {author} {\bibfnamefont {C.}~\bibnamefont {Kurdak}}, \bibinfo
  {author} {\bibfnamefont {D.-J.}\ \bibnamefont {Kim}}, \ and\ \bibinfo
  {author} {\bibfnamefont {Z.}~\bibnamefont {Fisk}},\ }\href@noop {} {\bibfield
   {journal} {\bibinfo  {journal} {arXiv preprint arXiv:1506.08233}\ }
  (\bibinfo {year} {2015}{\natexlab{b}})}\BibitemShut {NoStop}%
\bibitem [{\citenamefont {Rakoski}\ \emph {et~al.}(2017)\citenamefont
  {Rakoski}, \citenamefont {Eo}, \citenamefont {Sun},\ and\ \citenamefont
  {Kurdak}}]{PhysRevB.95.195133}%
  \BibitemOpen
  \bibfield  {author} {\bibinfo {author} {\bibfnamefont {A.}~\bibnamefont
  {Rakoski}}, \bibinfo {author} {\bibfnamefont {Y.~S.}\ \bibnamefont {Eo}},
  \bibinfo {author} {\bibfnamefont {K.}~\bibnamefont {Sun}}, \ and\ \bibinfo
  {author} {\bibfnamefont {i.~m.~c.}\ \bibnamefont {Kurdak}},\ }\href {\doibase
  10.1103/PhysRevB.95.195133} {\bibfield  {journal} {\bibinfo  {journal} {Phys.
  Rev. B}\ }\textbf {\bibinfo {volume} {95}},\ \bibinfo {pages} {195133}
  (\bibinfo {year} {2017})}\BibitemShut {NoStop}%
\bibitem [{\citenamefont {Hall}(1954)}]{hall1954variation}%
  \BibitemOpen
  \bibfield  {author} {\bibinfo {author} {\bibfnamefont {E.}~\bibnamefont
  {Hall}},\ }\href@noop {} {\bibfield  {journal} {\bibinfo  {journal} {Nature}\
  }\textbf {\bibinfo {volume} {173}},\ \bibinfo {pages} {948} (\bibinfo {year}
  {1954})}\BibitemShut {NoStop}%
\bibitem [{\citenamefont {Petch}(1953)}]{petch1953cleavage}%
  \BibitemOpen
  \bibfield  {author} {\bibinfo {author} {\bibfnamefont {N.}~\bibnamefont
  {Petch}},\ }\href@noop {} {\bibfield  {journal} {\bibinfo  {journal} {J. of
  the Iron and Steel Inst.}\ }\textbf {\bibinfo {volume} {174}},\ \bibinfo
  {pages} {25} (\bibinfo {year} {1953})}\BibitemShut {NoStop}%
\bibitem [{\citenamefont {Ellguth}\ \emph {et~al.}(2016)\citenamefont
  {Ellguth}, \citenamefont {Tusche}, \citenamefont {Iga},\ and\ \citenamefont
  {Suga}}]{ellguth2016momentum}%
  \BibitemOpen
  \bibfield  {author} {\bibinfo {author} {\bibfnamefont {M.}~\bibnamefont
  {Ellguth}}, \bibinfo {author} {\bibfnamefont {C.}~\bibnamefont {Tusche}},
  \bibinfo {author} {\bibfnamefont {F.}~\bibnamefont {Iga}}, \ and\ \bibinfo
  {author} {\bibfnamefont {S.}~\bibnamefont {Suga}},\ }\href@noop {} {\bibfield
   {journal} {\bibinfo  {journal} {Philosophical Magazine}\ }\textbf {\bibinfo
  {volume} {96}},\ \bibinfo {pages} {3284} (\bibinfo {year}
  {2016})}\BibitemShut {NoStop}%
\end{thebibliography}%

\end{document}



\title{Supplementary Material for Robustness of the Insulating Bulk in the Topological Kondo Insulator, SmB$_{6}$}

\author{Y. S. Eo}
	\email{eohyung@umich.edu}
	\affiliation{University of Michigan, Dept.~of Physics, Ann Arbor, Michigan 48109-1040, USA}
\author{A. Rakoski}
	\affiliation{University of Michigan, Dept.~of Physics, Ann Arbor, Michigan 48109-1040, USA}
\author{J. Lucien}
	\affiliation{University of Michigan, Dept.~of Physics, Ann Arbor, Michigan 48109-1040, USA}
\author{D. Mihaliov}
	\affiliation{University of Michigan, Dept.~of Physics, Ann Arbor, Michigan 48109-1040, USA}
\author{\c{C}. Kurdak}  
	\affiliation{University of Michigan, Dept.~of Physics, Ann Arbor, Michigan 48109-1040, USA}
\author{P. F. Rosa}
	\affiliation{Los Alamos National Laboratory}	
\author{D.-J. Kim}
	\affiliation{University of California at Irvine, Dept.~of Physics and Astronomy, Irvine, California 92697, USA}
\author{Z. Fisk}
	\affiliation{University of California at Irvine, Dept.~of Physics and Astronomy, Irvine, California 92697, USA}
 
\date{\today}

\maketitle

\subsection{Importance of Sample Preparation}
 
\begin{figure}[h]
\begin{center}
\includegraphics[scale=0.8]{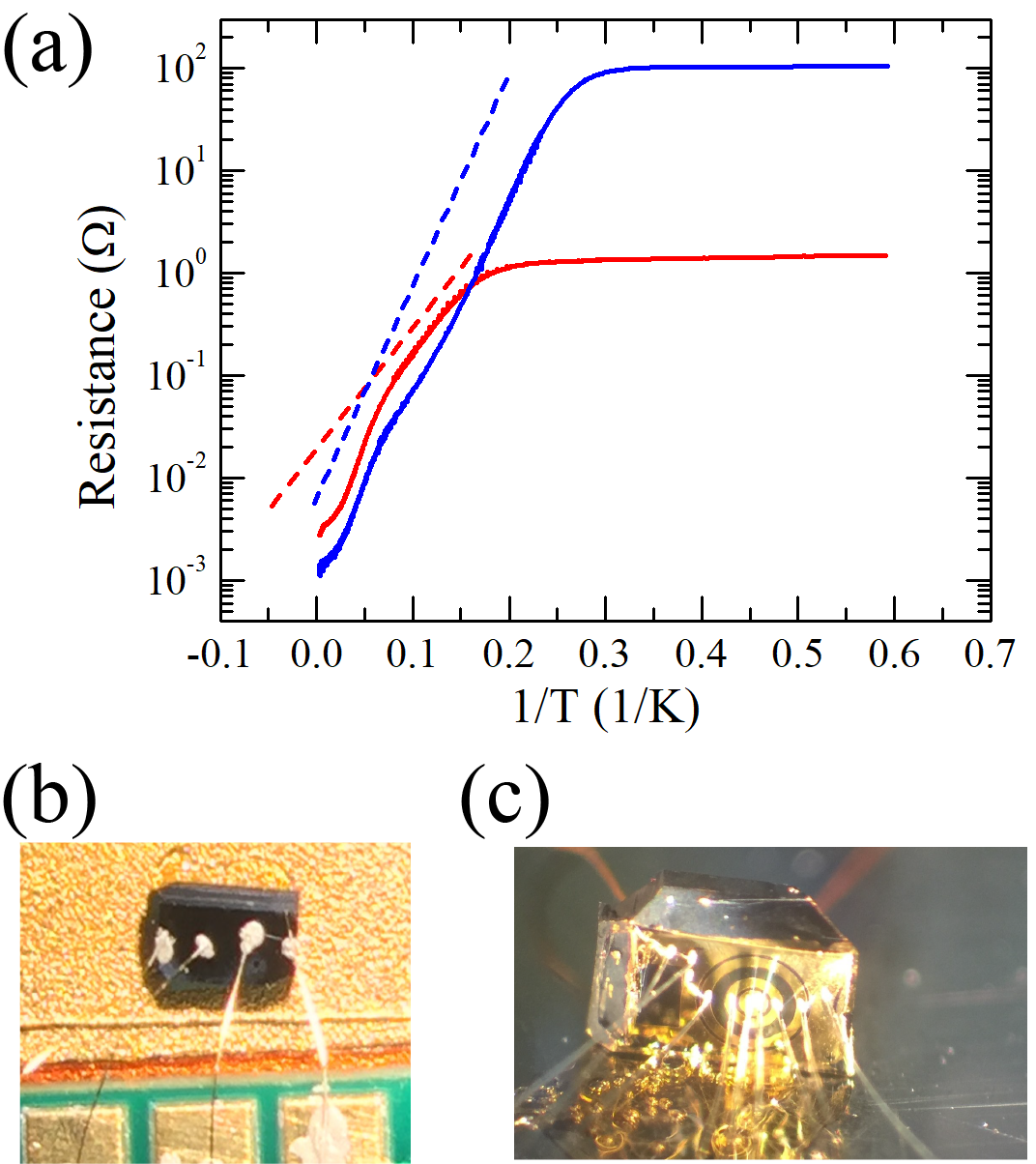}
\caption{(color) Comparison of transport results for different geometries on a sample grown from the batch of sample S2. (a) Resistance vs. $1/T$ of SmB$_{6}$ samples grown with 10 $\%$ less samarium from two different transport geometries. The solid blue curve is the data taken from the four-terminal Corbino disk, and the solid red curve is the data taken from a four-point contact on a raw sample (unpolished).  The dotted lines are examples of how a researcher might incorrectly fit the slope to analyze the bulk resistivity. (b) Picture of a four-point contact configuration on a raw sample. (c) Picture of a four-terminal Corbino disk. }
\label{Fig:S1}
\end{center}
\end{figure}

In Supplementary A, we show that choosing the proper transport geometry and preparing the sample surface is critically important for bulk studies of SmB$_{6}$ at low temperatures because current that flows on every surface nonuniformly can complicate the analysis. We have presented further details about geometry and surface preparation elsewhere\cite{wolgast2015magnetotransport}. In addition, we have recently noticed that cracks hidden underneath the surface may provide additional conduction paths \cite{wolgast2015conduction}. Fig.~\ref{Fig:S1} shows the two resistance curves as a function of inverse temperature. The solid red curve is measured from a standard 4-point contact, as shown in Fig~(\ref{Fig:S1})~(b), without any surface preparation. The solid blue curve is measured from a Corbino disk after carefully polishing the surface, as shown in Fig~(\ref{Fig:S1})~(c). The two samples are from the same crystal growth batch of sample S2. Consistent with our previous reports\cite{wolgast2015magnetotransport, wolgast2015conduction}, depending on the transport geometry and surface preparation, the resistance plateau at low temperatures changes greatly. The Corbino disk geometry, shown in a solid blue line, reaches up to about 100 $\Omega$, whereas the 4-point geometry, shown in a solid red curve, reaches up to about 1 $\Omega$. The samples that are prepared properly obtain a larger resistance plateau. 
 
 Now let us compare the resistance features of the two data sets in the bulk-dominated regime. The high-temperature ranges from 0 - 0.5 K$^{-1}$ seem to be parallel to each other, although the range is too small to be decisive. The two data sets shows the hump feature near 0.1 K$^{-1}$, which is predicted in the model by A. Rakoski et al.\cite{PhysRevB.95.195133}. However, the two sets of data have a slightly different magnitude and position. The resistance from the 4-point geometry (solid red curve) shows the hump feature at 10 - 12 K extends over a broader range. In fact, this feature is almost to the resistance plateau. Therefore, the four-point geometry does not provide a large enough range for extracting the bulk activation energy. In Fig~\ref{Fig:S1}~(a), we also show how one might fit the resistance data for extracting the bulk activation energy in dashed lines. The data from the four-point geometry shows that the extracted bulk activation energy can be estimated incorrectly. 
 
\subsection{Crystal characterization}
 In Supplementary B, we report on various experimental methods that we used for characterizing the crystals. In short, despite our effort using various methods to characterize the samples, we were not able to find the accurate (point and extended) defect concentration of each samples. Here, we use various methods such as X-ray diffraction, Auger electron spectroscopy, Vickers microhardness, and high resolution transmission electron microscopy.  

\subsubsection{Vikers Microhardness}

During sample preparation, we noticed that non-stoichiometrically grown SmB$_{6}$ samples polish significantly faster than pure samples. To check if this observation was consequential, we tested the microhardness on the samples that were used in the transport studies. It is known that the introduction of defects into a crystalline material can influence the microhardness. The relationship between hardness and point and extended defects is not highly universal. However, one noticeable relation, although many exceptions exist, is the Hall-Petch relation, which states that the hardness is inversally proportional to the square root of the size of grains\cite{hall1954variation,petch1953cleavage}, which is an extended defect.  

\begin{figure}[h]
\begin{center}
\includegraphics[scale=0.8]{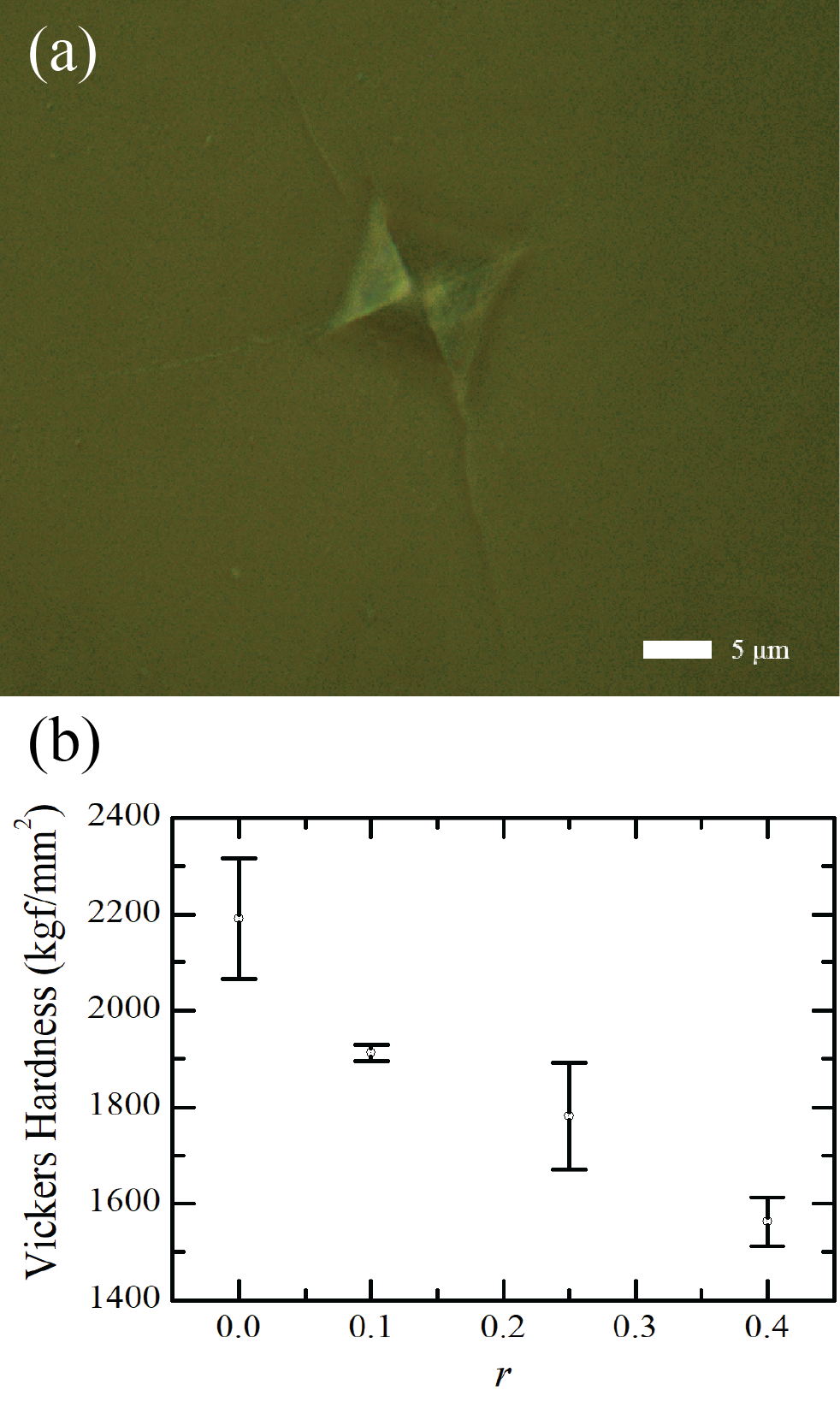}
\caption{(color) (a) Example of an indentation after hardness measurement. (b) Vickers microhardness data of S1, S2, S3, and S4.}
\label{Fig:S2}
\end{center}
\end{figure}

Microhardness of the samples was measured using the Vickers hardness test, which is performed by applying a force on the flat (001) surface using a pyramidal diamond indenter and then measuring the area of indentation. A picture of a typical indentation is shown in Fig~\ref{Fig:S2}~(a). For each specimen, we typically used multiple indentations made with a force of 300 kgf and measured the area in an optical microscope. This was repeated on multiple samples from the same batch to determine an average Vickers Pyramid Number (HV). The HVs of samples from the same batches as S1, S2, S3, and S4 are shown in Fig~\ref{Fig:S2}~(b), and also given in Table I of the main text. The Vikers microhardness shows that all four samples indeed have a different physical property.

\subsubsection{Auger Spectroscopy}

\begin{figure}[h]
\begin{center}
\includegraphics[scale=0.9]{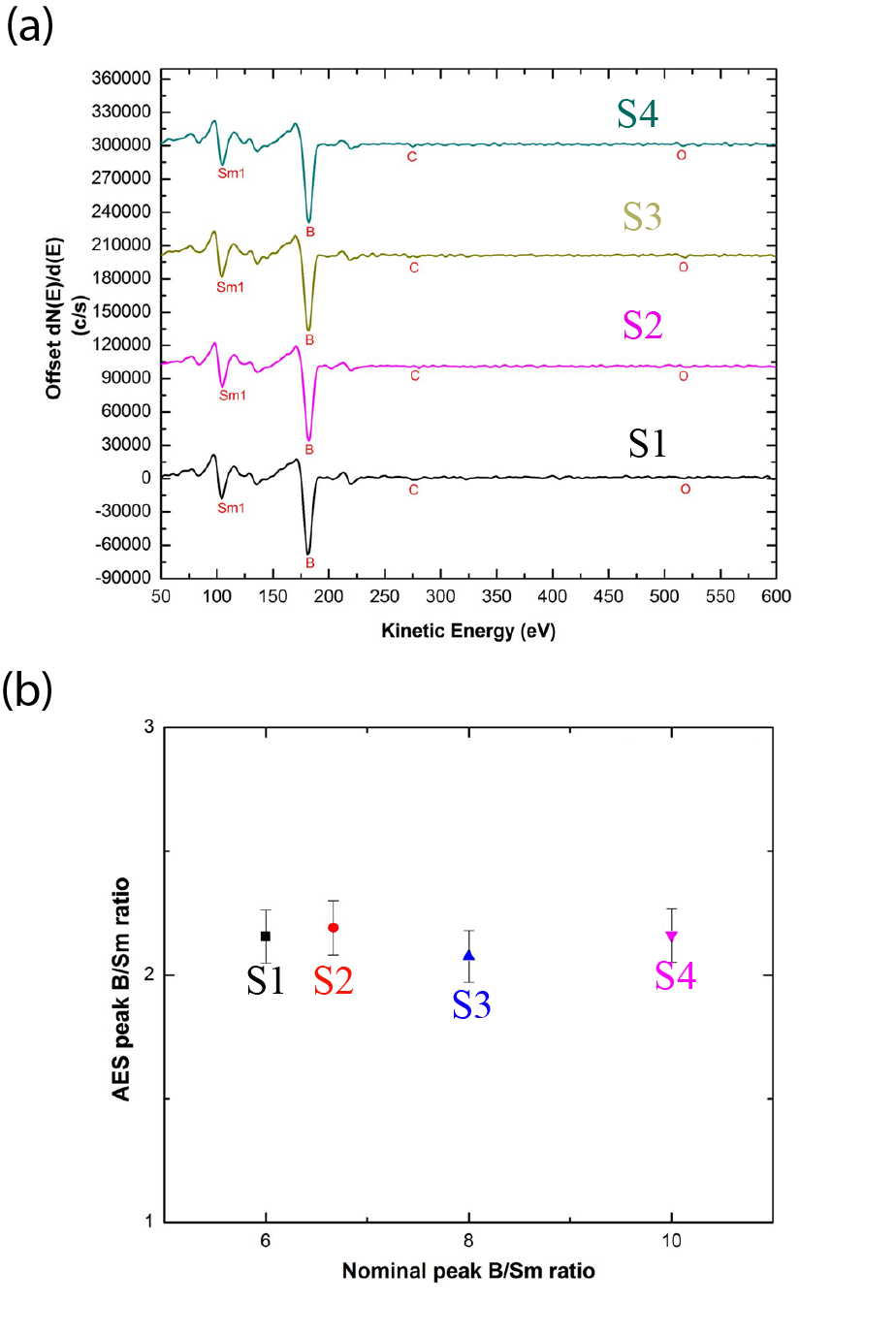}
\caption{(color) Auger electron spectroscopy results. (a) Auger electron spectroscopy results of the samples S1, S2, S3, and S4 batches. (c) Estimated (nominal) B/Sm ratio.}
\label{Fig:S3}
\end{center}
\end{figure}

We used Auger Electron Spectroscopy (AES) in order to find the boron-to-samarium ratio of each of the samples. The elements can be identified by the peak positions of the energies, and the peak intensities can be used for obtaining the concentration of the elements. It is well known that the surface of SmB$_{6}$ samples easily oxidize, and the oxygen removal is very challenging even at high vacuum conditions\cite{ellguth2016momentum}. All four samples were cleaned at a spot on the surface using Ar ion sputtering, already used by a previous study\cite{ellguth2016momentum}, before collecting three spectra in succession. We followed the same procedure for all four samples for consistency. We still notice traces of carbon and oxygen on the cleaned surface, even under ultrahigh vacuum ($\sim 10^{-10}$ Torr). Fig.~\ref{Fig:S3}~(a) displays spectra from all four samples after removing significant amount of carbon and oxygen signals after sputtering. By comparing the relative peak heights for Sm and B and neglecting the small contribution from the C and O peaks, the B/Sm ratio was found. From the Auger spectroscopy, we do not see a significant difference in the B/Sm ratio. In Fig.~\ref{Fig:S3}~(b) shows the boron-to-samarium ratio from the peaks. We note that this ratio estimation does not take into account all the prefactors, so the reader should not take the absolute magnitude seriously. Here, we see that all four samples from different growths do not show a significant boron-to-samarium ratio difference. 

\subsubsection{X-ray Diffraction}
\begin{figure}[h]
\begin{center}
\includegraphics[scale=0.8]{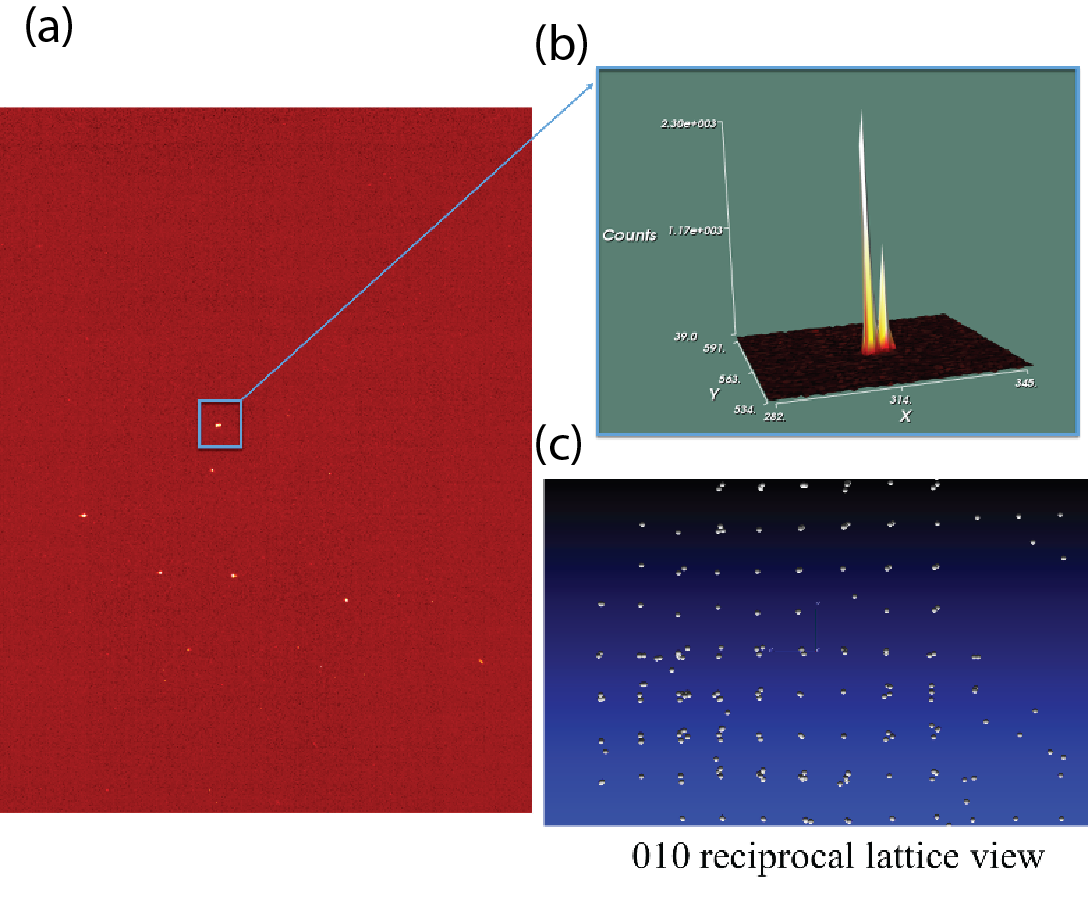}
\caption{(color) Example of twinning signatures in sample S3 (a) Representative frame. (b) Twinning peaks. (c) (010) reciprocal lattice view.}
\label{Fig:S4}
\end{center}
\end{figure}
In the main text, one suspicion about the origin of the residual bulk conduction in the non-stoichiometrically-grown SmB$_6$ samples was that extended defects exist in the bulk, extending globally in the sample. From the diffractometer (Bruker D8) several crystals presented signs of twinning, i.e., the presence of two or more crystals of the same species joined together in different orientations. An example of such behavior is shown in Figure~\ref{Fig:S4} for a small crystal dimensions $30$ x $50$ x $60$ $\mu$m$^{3}$ that was detached from sample S3. Fig.~\ref{Fig:S4}~(a) shows a representative frame in which some reflections are sharp whereas others split. Further, Fig.~\ref{Fig:S4}~(c) shows the presence of clusters at several lattice points in the (001) reciprocal space view. Nevertheless, it is possible to fit the data using a least-square procedure to the cubic space group $Pm-\bar{3}m$, and $93\%$ of the reflections can be assigned to a single domain. About $3\%$ of the reflections, however, belong to another domain rotated by 174 degrees from the main one. Another sign of twinning comes from the $E^{2}-1$ statistics of the diffracted data, where $E$ is the normalized structure factor. Our data show that the mean variance $\langle | E^{2} - 1 | \rangle$ is much lower than the expected value of $1$ for a centrosymmetric structure. This decrease is a well-known indication of twinning because the overlap of the twinned diffraction patterns tends to average out intensities as strong and weak reflections sometimes overlap.

\subsubsection{Transmission Electron Microscopy}
\begin{figure}[h]
\begin{center}
\includegraphics[scale=0.8]{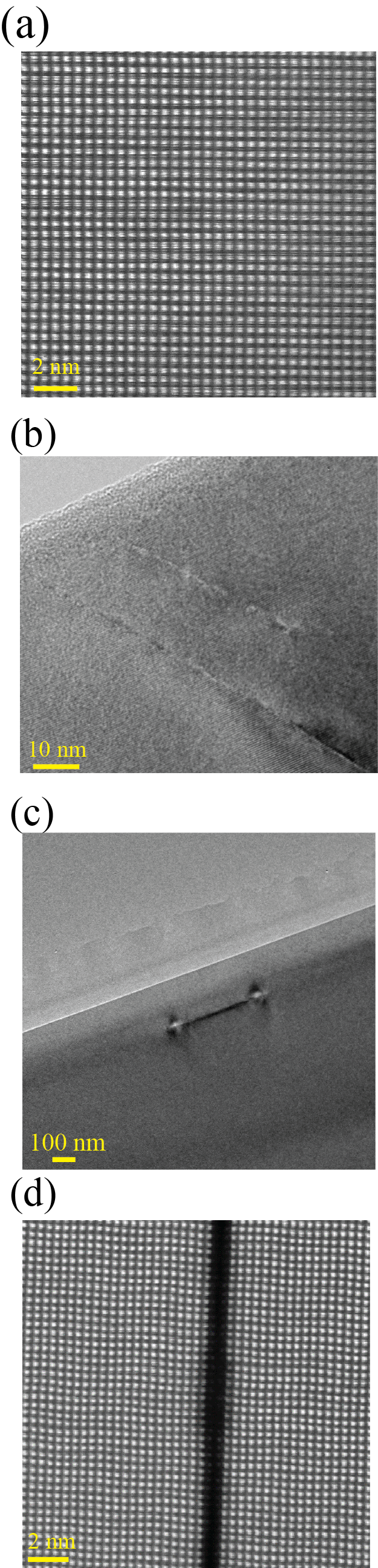}
\caption{TEM Image of sample S4. (a) Image of specimen from the active region of sample S4. The white dots are Sm atoms. (b) Line-shaped nanocrack observed near the surface. (c) Dumbbell-shaped nanocrack near the surface. (d) A line of missing layers near the surface region.}
\label{Fig:S5}
\end{center}
\end{figure}

 To search for extended defects, we used transmission electron microscopy (TEM). In order for the extended defects to explain the low-temperature bulk resistivity plateaus or saturation of the inverted resistance measurements in the main text, they must extend throughout the bulk (top-to-bottom surface). In order to find the most dramatic effect possible, we have tried to probe through thin specimens that were cut from the active transport region of sample S4, which shows the loudest plateau feature in bulk resistivity. Interestingly, we have found small imperfections near the surface, as shown in Fig.~\ref{Fig:S5} However, these features only exist near the surface of the active region of the Corbino disk, and therefore are not responsible for the bulk resistivity. Also, the position and the length scales of these features are in the order of 0.1 $\mathrm{\mu}m$. Interestingly, this magnitude is of the same order of magnitude of the Al$_{2}$O$_{3}$ particles (0.3 $\mathrm{\mu}m$) that were used in our final polishing procedure. This suggests that, most likely, these features are created during the surface preparation rather than extended defects created during the crystal growth. Also, these features may influence the surface transport studies, but should not influence the bulk resistivity since they do not have length scales of the thickness of the sample (several hundred microns). Further investigation is needed if conduction through 1D (or higher dimensional) defects are the very cause of the bulk resistivity plateaus.

\subsection{Converting Measured Resistances to Bulk Resistivity}
\begin{figure}[h]
\begin{center}
\includegraphics[scale=1]{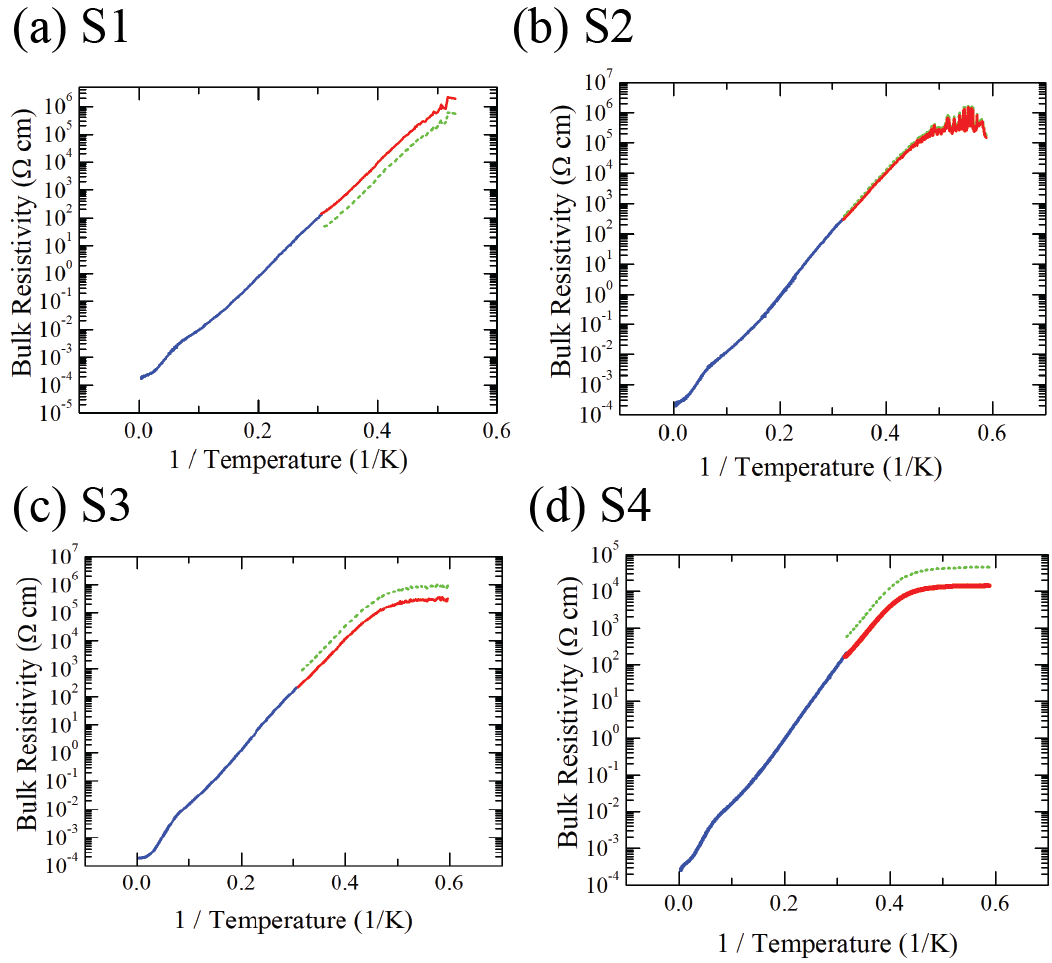}
\caption{(color) Patching the low temperature and high temperature regime. The blue curve is the high temperature regime extracted from Eq.~\ref{Eq:twochannel}. The green dotted line is extracted from the Eq.~\ref{Eq:Inverted}. The red curve is adjusted to match the high temperature data. (a) Sample S1. (b) Sample S2. (c) Sample S3. (d) Sample S4.}
\label{Fig:S6}
\end{center}
\end{figure}

 In Supplementary C, we discuss how the bulk resistivity ($\rho_{b}$) was found from the resistance measurements ($R_{Std}$ and $R_{Inv}$). As mentioned in the main text, the two-channel model is a good approximation for the $R_{Std}$ in both the low temperature (surface-dominated) and the high temperature (bulk-dominated) regime: 
\begin{equation}
	R_{Std} = C_0 (\rho_{s}^{-1} + \gamma \rho_{b}^{-1})^{-1}.
	\label{Eq:twochannel}
\end{equation}
 In the surface-dominated regime, $\rho_b / \rho_{s} t \rightarrow 0$, the equation reduces to:
\begin{equation}
	R_{Std} = C_{0}\rho_{s},
	\label{Eq:twochannel_surface}
\end{equation}
where for a Corbino disk geometry, $C_0$ is ln$(r_{out}/r_{in})/2\pi$. In the bulk dominated regime, $\rho_b / \rho_{s} t \rightarrow \infty$, we can express Eq.~(\ref{Eq:twochannel}) as: 
\begin{equation}
	R_{Std} = \frac{C_{0}\rho_{b}}{\gamma}.
	\label{Eq:twochannel_bulk}
\end{equation}
Alternatively, we can express as:
\begin{equation}
	R_{Std} = \frac{C_{-1}\rho_{b}}{t}.
	\label{Eq:twochannel_bulk2}
\end{equation}
The inverted resistance is again:
\begin{equation}
	R_{Inv} = C_{1} t \frac{\rho_{s}^2}{\rho_{b}},
	\label{Eq:Inverted}
\end{equation}
where we have defined $C_{-1}=(t/\gamma)C_{0}$, for convenience. From Eq.~(\ref{Eq:twochannel}) and Eq.~(\ref{Eq:Inverted}), $\rho_{b}$ can be found in the full range if we know $C_{-1}$, $C_{0}$, and $C_{1}$. When the Corbino disk dimensions are fixed, these coefficients are a function of the sample's thickness. We used finite element analysis from Comsol Multiphysics AC/DC module to find these coefficients. The values for the corresponding thickness is shown in Table~\ref{Tab:summary}. The results are shown in Fig.~\ref{Fig:S6}. Typically, the bulk resistivity found from the high temperature regime and the low temperature regime mismatched by a factors of 3-4. We have previously reported that this is most likely due to the imperfection of the transport geometry such as the misalignment of the two Corbino disks, or the inhomogeneous surface quality (e.g. differences in the top and bottom surfaces). We also note that aluminum flux trapped in the bulk cannot explain the low temperature resistivity plateau. This will only change the values of the coefficients, $C_{-1}$ and $C_{1}$. An aluminum flux path that intersects both the top and bottom surfaces may short the current path. To avoid this effect, we have polished our samples on both surfaces that do not show any aluminum fluxes on the surface. 
\begin{table}[h]
\squeezetable
\centering 
\begin{tabular}{c c c c c} 
\hline\hline 
Sample & Thickness ($\mu$m) & C$_{1}$ & C$_{-1}$ \\ [1ex]
\hline 
S1* & 310 & 1.80 $\times$ 10$^{-4}$& 0.18  \\ 
S2 & 440 & 1.16 $\times$ 10$^{-3}$& 0.26  \\
S3 & 250 & 8.70 $\times$ 10$^{-3}$& 0.15  \\
S4 & 150 & 3.06 $\times$ 10$^{-2}$& 0.10 \\[1ex] 
\hline 
\end{tabular}
\caption{Summary of the geometric prefactors that were found numerically for the corresponding thickness of the samples. *The inverted resistance for S1 was found from a single surface Corbino disk measurement.}
\label{Tab:summary}
\end{table}

\subsection{Considering Aluminum Inclusions for Bulk Transport}

\begin{figure}[t]
\begin{center}
\includegraphics[scale=0.6]{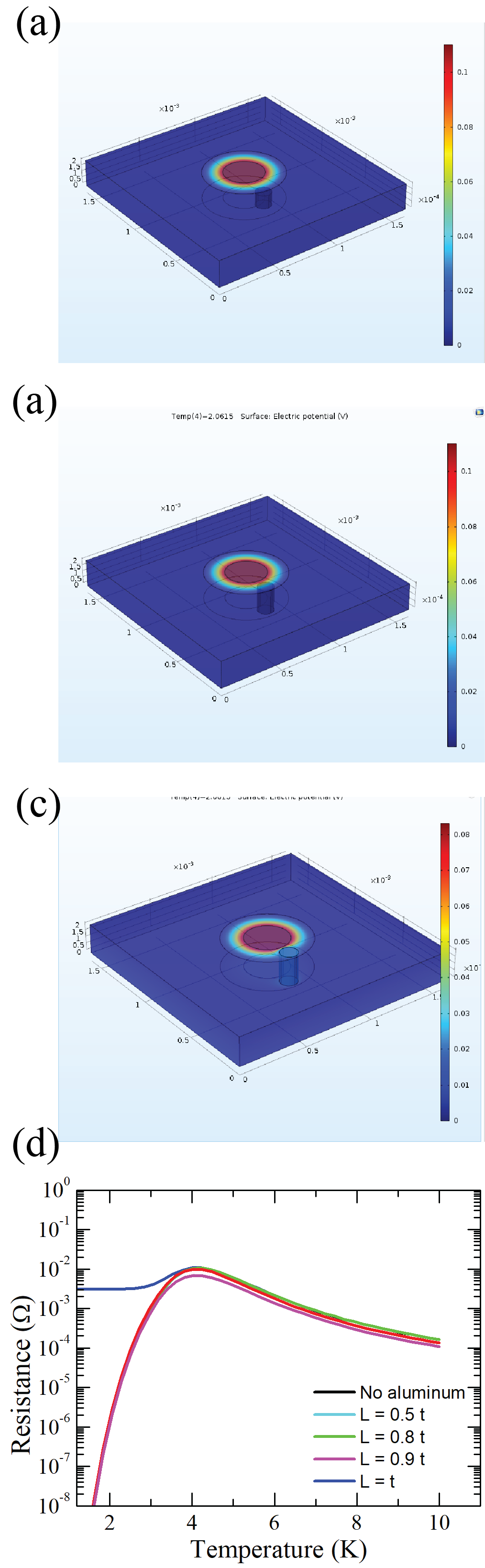}
\caption{(color) Demonstration of a highly conducting cylinder located in the bulk. (a) The conducting cylinder that extends 50 percent of the thickness of the sample. (b) The conducting cylinder that extends 90 percent of the thickness of the sample. (c) The conducting cylinder that touches the top and bottom surfaces.}
\label{Fig:S8}
\end{center}
\end{figure}

 In this section, we study the role of aluminum inclusions by numerical simulations. One question the reader may have is that the residual bulk resistivity plateau at low temperatures may be due to aluminum inclusions. Here, we numerically demonstrate a double-sided Corbino disk, and see how the current behaves when a highly conducting cylinder is present in the bulk. Example demonstrations are shown in Fig.~\ref{Fig:S8}~(a) to (c). At low temperatures, below the bulk-to-surface crossover, when the inclusion is trapped inside the bulk, as shown in Fig.~\ref{Fig:S8}~(a) to (b), the qualitative bulk behavior is identical without any inclusions even when the conducting cylinder extends 90 perecent of the total thickness. However, when the cylinder touches both the top and bottom surface, as shown in Fig.~\ref{Fig:S8}~(c), the current flows from the top-to-bottom surface. 
 
 The simulated inverted resistance values are shown in Fig.~\ref{Fig:S8}~(d). When the aluminum inclusions are trapped inside the bulk, the inverted resistance behaves identically with only a change in the geometric prefactor, $C_{1}$. Only the case when the inclusion touches the surfaces shows a high resistance plateau. 
 
 In conclusion, a qualitatively similar bulk resistivity plateau can appear if aluminum is shorting the sample on both surfaces. Of course, we have polished all of our samples that no aluminum inclusions can be seen in both of the surfaces through a polarizable microscope. Furthermore, our samples were sufficiently etched with acid (10 - 20 percent hydrochloric acid) after fine polishing and also after patterning the gold electrodes, making sure no aluminum bubbling off from the surface can be seen. Since it is very well known that large inclusions (hundreds of micron size) in SmB$_{6}$ can be easily etched away with dilute hydrochloric acid, we believe any small aluminum inclusions that are exposed but cannot be seen through the microscope would be very unlikely to be present.

\bibliography{RobustBulkSmB6_Supplementary}